\documentclass[aps,prd,preprint,groupedaddress,showpacs]{revtex4}

\usepackage{graphicx}
\usepackage{color}

\newcommand{\bfr}{\mbox{\boldmath $r$}}
\newcommand{\bfx}{\mbox{\boldmath $x$}}

\begin{document}


\preprint{ICRR-Report-578-2010-11}
\preprint{IPMU 11-007}

\title{Destruction of $^7$Be in big bang nucleosynthesis via long-lived
  sub-strongly interacting massive particles as a solution to the Li problem}


\author{Masahiro Kawasaki$^{1,2}$ and Motohiko Kusakabe$^{1}$\footnote{kusakabe@icrr.u-tokyo.ac.jp}}

\affiliation{
$^1$Institute for Cosmic Ray Research, University of Tokyo, Kashiwa,
Chiba 277-8582, Japan\\
$^2$Institute for the Physics and Mathematics of the Universe,
University of Tokyo, Kashiwa, Chiba 277-8582, Japan}


\date{\today}

\begin{abstract}

We identify reactions which destroy $^7$Be and $^7$Li during big bang
 nucleosynthesis (BBN) in the scenario of BBN catalyzed by a long-lived
 sub-strongly interacting massive particle (sub-SIMP or $X$ particle).
 The destruction associated with non radiative $X$ captures of the nuclei
 can be realized only if the interaction strength
 between an $X$ particle and a nucleon is properly weaker than that between two
 nucleons to a degree depending on the mass of $X$.  Binding energies of nuclei to an $X$ particle are estimated taking the mass and the interaction strength to nuclei of
 the $X$ as input parameters.  Nuclear reaction rates
 associated with the $X$ are estimated naively, and adopted in
 calculating evolutions of nuclear abundances.  We suggest that
 the $^7$Li
 problem, which might be associated with as-yet-unrecognized particle processes
 operating during BBN, can be solved if the $X$ particle
 interacts with nuclei strongly enough to drive $^7$Be destruction but
 not strongly enough to form a bound state with $^4$He of relative
 angular momentum $L=1$.  Justifications of this
 scenario by rigorous calculations of reaction rates using quantum
 mechanical many-body models are highly desirable since this result
 involves many significant uncertainties.

\end{abstract}



\maketitle

\section{Introduction}\label{sec1}

The standard big bang nucleosynthesis (BBN) model predicts primordial light
element abundances which are more or less consistent with abundances
inferred from observations of old distant astronomical objects.
Deviations from the standard BBN (SBBN) model are, therefore, constrained if
predicted abundances in theoretical models change from those in the SBBN.  Constraints on the existence of long-lived exotic particles which
interact with nuclei by strong force~\cite{Dicus1980,Plaga1995,Mohapatra:1998nd,Kusakabe:2009jt} or Coulomb force~\cite{Pospelov:2006sc,Kohri:2006cn,Cyburt:2006uv,Hamaguchi:2007mp,Bird:2007ge,Kusakabe:2007fu,Kusakabe:2007fv,Kusakabe:2010cb,Jedamzik:2007cp,Jedamzik:2007qk,Kamimura:2008fx,Pospelov:2007js,Kawasaki:2007xb,Jittoh:2007fr,Jittoh:2008eq,Jittoh:2010wh,Pospelov:2008ta,Khlopov:2007ic}
have been derived as well as those on the decay of long-lived exotic
particles into standard model particles which have electromagnetic or
hadronic
interactions~\cite{Ellis:1984er,Cyburt:2002uv,Ellis:2005ii,Terasawa:1988my,Kawasaki:1993gz,Kawasaki:1994af,Holtmann:1996cq,Kawasaki:2000qr,Kawasaki:2004yh,Kawasaki:2004qu,Kanzaki:2006hm,Kawasaki:2008qe,Cumberbatch:2007me,Reno:1987qw,Dimopoulos:1987fz,Dimopoulos:1988zz,Dimopoulos:1988ue,Khlopov:1993ye,Sedelnikov:1987ef,Jedamzik:1999di,Jedamzik:2004er,Jedamzik:2004ip,Jedamzik:2005dh,Jedamzik:2006xz,Kusakabe:2006hc,Kusakabe:2008kf,Pospelov:2010cw,Pospelov:2010kq}.

A prominent problem relating to the abundances predicted in the SBBN model and inferred from observations is lithium
problem~\cite{Melendez:2004ni,Asplund:2005yt}.  Primordial lithium abundances
are inferred from measurements in metal-poor halo stars (MPHSs).  Observed
abundances are roughly constant as a function of
metallicity~\cite{Spite:1982dd,Ryan2000,Melendez:2004ni,Asplund:2005yt,bon2007,Shi:2006zz,Aoki:2009ce}
at $^7$Li/H$=(1-2) \times 10^{-10}$.  The theoretical prediction by the
SBBN model is, however, a factor of
2--4 higher, e.g., $^7$Li/H=$(5.24^{+0.71}_{-0.67})\times
10^{-10}$~\cite{Cyburt:2008kw}, when its only parameter, the baryon-to-photon
ratio, is deduced from the observation with Wilkinson
Microwave Anisotropy Probe (WMAP) of the cosmic microwave background (CMB)
radiation~\cite{Larson:2010gs}.  This discrepancy indicates some
mechanism of $^7$Li reduction having operated in some epoch from the BBN
to this day.  One possible astrophysical process to reduce $^7$Li
abundances in stellar surfaces is the gravitational settling in the
model including a combination of the atomic and turbulent
diffusion~\cite{Richard:2004pj,Korn:2006tv}.  The precise trend of Li
abundance as a function of effective temperature of stars in the metal-poor
globular cluster NGC 6397 is, however, not reproduced
theoretically~\cite{Lind:2009ta}.  

$^6$Li/$^7$Li isotopic ratios of MPHSs
have also been measured spectroscopically.  The $^6$Li abundance as high
as $^6$Li/H$\sim 6\times10^{-12}$ was suggested~\cite{Asplund:2005yt},
which is about 1000 times higher than the SBBN
prediction~\footnote{Recently, a new measurement of the cross section of
radiative $\alpha$ capture by deuteron and the $^6$Li abundance
predicted based upon the result have been reported.~\cite{Hammache:2010bt}}.  Convective
motions in the atmospheres of MPHSs could cause systematic asymmetries in
the observed line profiles and mimic the presence of
$^6$Li~\cite{Cayrel:2007te}.  A few or
several MPHSs, however, have high $^6$Li abundances larger than levels caused
by this effect~\cite{Steffen:2009yr}.   This high $^6$Li
abundance is a problem since the standard Galactic cosmic ray
nucleosynthesis models predict negligible amounts of $^6$Li yields
compared to the observed level in the epoch corresponding to the
metallicity of the stars, i.e., [Fe/H]
$<-2$~\cite{Prantzos2006}~\footnote{[A/H]=log(A/H)$-$log(A/H)$_\odot$ is
the number ratio of nuclide $A$ to H measured in a logarithmic scale
normalized to the solar value.}.

The possibility that
the $^7$Li and $^6$Li problems stem from uncertainties in nuclear
reactions used in theoretical BBN calculation is unlikely~\cite{Boyd:2010kj} unless there remain to be observed new
resonant states contributing to $^7$Be
destruction~\cite{Cyburt:2009cf,Chakraborty:2010zj}.  The $^7$Li
reduction needs a destruction mechanism of $^7$Be during or after BBN
and before stellar activities since the $^7$Li nuclei observed in MPHSs
are thought to have originated from the electron capture process of
$^7$Be which is produced in the BBN.

Some particle models beyond the standard model include heavy ($m \gg 1$~GeV) long-lived colored
particles.  The scenarios, i.e., split
supersymmetry~\cite{ArkaniHamed:2004fb,ArkaniHamed:2004yi}, weak scale
supersymmetry with a long-lived gluino~\cite{Raby:1997bpa,Shirai:2010rr,Covi:2010au} or
squark~\cite{Sarid:1999zx} as the next-to-lightest supersymmetric
particles, and extended theories with new kinds of colored
particles~\cite{Hisano:2010bx,Nakayama:2010vs}, may be tested in
experiments such as the Large Hadron Collider.
The heavy colored particles would be confined at temperatures below the deconfinement
temperature $T_C\sim 180$~MeV inside exotic heavy hadrons, i.e., strongly
interacting massive particles (SIMPs) which we call $X$
particles~\cite{Kang:2006yd}.  Their thermal relic
abundances after the freeze-out of annihilations depend on the
annihilation cross sections, and theoretical estimates predict various
values which extend over more than several order of magnitude at the heavy
mass limit~\cite{Baer:1998pg}.  

If the annihilation cross section is not
different from a typical value for strong interaction, i.e.,
$\sigma\sim
\mathcal{O}({\rm GeV}^{-1})^2$, however, the final abundance of $X$ particles
can be derived under the assumption that their abundances are fixed when
the annihilation rate becomes smaller than the Hubble expansion rate of
the universe~\cite{Kang:2006yd}.  The relic abundance can then be written
\begin{equation}
\frac{N_X}{s} \sim \sqrt[]{\mathstrut \frac{15}{\pi}}
 \frac{g_\ast^{1/2}}{g_{\ast s}} \frac{m^{1/2}}{\sigma T_B^{3/2} m_{\rm Pl}}~~,
\label{eq1}
\end{equation}
where
$N_X$ is the number density of the $X$ particle,
$s=2\pi^2 g_{\ast s} T^3/45$ is the entropy density with $g_{\ast s}\sim 10$ the total number of
effective massless degrees of freedom in terms of entropy~\cite{kolb1990} just below the QCD
phase transition,
$g_\ast$ is the total number of effective massless degrees of
freedom in terms of number~\cite{kolb1990},
$m$ is the mass ($m\gg 1$~GeV) of the heavy long-lived colored
particles,
$\sigma$ is the annihilation cross section of the $X$ particle,
$T_B$ is the temperature of the universe at which the $X$-particles are
formed,
and $m_{\rm Pl}$ is the Planck mass.
The number abundance of the $X$s with respect to that of baryons is then
\begin{equation}
\frac{N_X}{n_b} \sim 0.5\times 10^{-8} \left(\frac{g_\ast}{10.75}\right)^{1/2}
 \left(\frac{m}{\rm TeV}\right)^{1/2} \left(\frac{T_B}{180 {\rm
    MeV}}\right)^{-3/2} \left(\frac{\sigma}{m_\pi^{-2}}\right)^{-1}~~,
\label{eq2}
\end{equation}
where $n_b$ is the number density of baryons, and
$m_\pi\sim 140$~MeV is the mass of pion.  The thermal relic
abundance is inversely proportional to the annihilation cross section
which depends on the particle theory.  In addition, there might be a
nonthermal production of long-lived colored particles which is not
directly related to the thermal production.  The final abundance of the
$X$ is, therefore, uncertain.  So we consider the $X$
abundance as a free parameter in this paper.

Observational constraints on hypothetical SIMPs have been studied~\cite{Wolfram:1978gp,Dover:1979sn,Starkman:1990nj}.  Effects of
exotic neutral stable hadrons on BBN were studied in Ref.~\cite{Dicus1980}.
The authors assumed that the strong force between a nucleon and a exotic hadron
($X$) is
similar to that between a nucleon $N$ and a $\Lambda$ hyperon.  In addition,
new hadrons are assumed to be captured in a bound state of $^4$He plus $X$ after BBN.
Based upon an analytic estimation, they suggested that beryllium has
the largest number fraction $A_X/A$ of bound states with the hadrons among the light
elements produced in BBN, where the $A$ and $A_X$ are a nuclide $A$ and a bound state of
$A$ with a hadron $X$.  Mohapatra and Teplitz~\cite{Mohapatra:1998nd}
estimated the cross section of $X$ capture by $^4$He, and suggested that a large fraction
of free $X$ particles would not become bound into light nuclides and
remain free contrary to the previous suggestion~\cite{Dicus1980}.  In
deriving the result of those two studies, it has been assumed that exotic hadrons interact with normal nuclei by
typical strengths of strong interaction and implicitly assumed that its mass is about that
of $\Lambda$ hyperon, i.e., $m_X \sim 1.116$~GeV~\cite{Particle2010}.

Effects on BBN of long-lived exotic hadrons of $m\gg 1$~GeV have been
studied recently~\cite{Kusakabe:2009jt}.  The authors have assumed that the
interaction strength between an $X$ particle and a nucleon is similar to
that between nucleons.  Rates of many reactions associated
with the $X$ particle were estimated, and a network calculation of the
nucleosynthesis including effects of the $X$ was performed.  The constraint on the
decay lifetime of such $X$ particles, i.e., $\tau_X \lesssim 200$~s
was derived from a comparison of calculated abundances with
observational abundance constraints of light elements.

Two interesting predictions of the model~\cite{Kusakabe:2009jt} is
signatures of the $X$ particles on primordial abundances which should be seen in future astronomical observations:
1) $^9$Be and B can be produced in amounts more than predicted in the SBBN.  Future
observations of Be and B abundances in MPHSs may show primordial
constant values originating from the BBN catalyzed by the $X$ particle.
2) The isotopic ratio $^{10}$B/$^{11}$B tends to be very high.  This is
different from predictions of other models for boron production, i.e., the
cosmic ray nucleosynthesis ($^{10}$B/$^{11}$B$\sim
0.4$~\cite{Prantzos1993,Ramaty1997,Kusakabe2008}) or the supernova
neutrino process ($^{10}$B/$^{11}$B$\ll
1$~\cite{Woosley:1995ip,Yoshida:2005uy}). They concluded that the $^6$Li or $^7$Li problems is not solved under
their assumption.

Since interactions between long-lived exotic hadrons $X$ and a nucleon are not
known as well as their masses, we are investigating effects of such
particles in various cases of interaction strengths and masses.  We
found on the way a new possibility that reactions associated with the $X$
particle reduce $^7$Be abundance and that the $^7$Li problem is solved.

In this paper we report details of the destruction mechanism of $^7$Be
in the presence of the $X$ particle.  We carry out a network calculation of BBN in the presence
of a long-lived SIMP $X^0$ of a zero charge taking the mass and the
strength of interaction with a nucleon as characterizing parameters.  In Sec.~\ref{sec2} assumptions on the $X^0$ particle, estimations for
binding energies between nuclei and an $X^0$, and rates of important
reactions are described.  Effects of the $X^0$ decay inside
$X$-nuclei are not considered in our model.  They should be addressed in
the future.  In Sec.~\ref{sec3} the destruction processes of $^7$Be
and $^7$Li are identified.  With results of the network calculations of
BBN, we delineate the parameter region in which the $^7$Be and $^7$Li
destructions possibly operate.  If the $X^0$ particle
interacts with nuclei strongly enough to drive $^7$Be destruction but
not strongly enough to form a bound state with $^4$He of relative
angular momentum $L=1$, then it might solve the $^7$Li problem of
standard BBN.  In Sec~\ref{sec4} conclusions of
this work are summarized, and this model for $^7$Li reduction is
compared with other models.

\section{Model}\label{sec2}

A strongly-interacting massive particle (SIMP) $X$ of spin zero and charge zero
is assumed to exist during the epoch of BBN.  Its mass is one parameter
since it is not known a priori at the moment.  Two types of nuclear potentials
between an $X^0$ and a nucleon ($XN$) are considered in this study.  One is the
Gaussian type given by
\begin{equation}
 v(r)=v_0 \delta \exp\left[-(r/r_0)^2\right],
\label{eq3}
\end{equation}
where
$v_0=-72.15$~MeV and $r_0=1.484$~fm~\footnote{These parameters have been
adjusted to fit the deuteron binding energy and low-energy triplet-even proton-neutron scattering phase shifts~\cite{Yahiro1986,Austern:1087zz}.}, and
the interaction strength is varied by changing $\delta$, the second
parameter.  When the $\delta$ equals unity then the binding energy of
deuteron, i.e., 2.224~MeV is obtained.  

The potential between an $X^0$
and a nuclide $A$ ($XA$) is given by
\begin{equation}
 V(r)=\int v(\bfx) \rho(\bfr^\prime)~d\bfr^\prime,
\label{eq4}
\end{equation}
where
$\bfr$ is the radius from an $X$ to the center of mass of $A$,
$\bfr^\prime$ is the distance between the center of mass of $A$ and a nucleon inside the nuclide $A$,
$\bfx=\bfr+\bfr^\prime$ is the distance between the $X$ and the nucleon, and
$\rho(\bfr^\prime)$ is the nucleon density of the nucleus
which is generally distorted by potential of an $X^0$ from
the density of normal nucleus.  Under the assumption of spherical symmetry in nucleon density, i.e., $\rho(r)$, the potential is written in the form of
\begin{eqnarray}
 V(r)&=&\pi v_0 \delta \frac{r_0^2}{r} \int_0^\infty dr^\prime r^\prime
 \rho(r^\prime) \nonumber\\
 & &\times \left\{\exp\left[-\frac{(r-r^\prime)^2}{r_0^2}\right]-\exp\left[-\frac{(r+r^\prime)^2}{r_0^2}\right]\right\}.
\label{eq5}
\end{eqnarray}

Another potential is a well given by
\begin{equation}
 v_{\rm w}(r)=v_{0{\rm w}}\delta_{\rm w}~~~~~({\rm for}~0\leq r < r_{0{\rm w}}),
\label{eq6}
\end{equation}
and $v_{\rm w}(r_{0{\rm w}} \leq r)=0$.  Parameters are fixed to be
$v_{0{\rm w}}=-20.06$~MeV and $r_{0{\rm w}}=2.5$~fm.  In order to make a comparison
between the two potential cases easy, integrals, i.e., $I=\int v(r) d\bfr$ for
both cases are made equal when $\delta=\delta_{\rm w}=1$.  This integral
is a characteristic quantity which is related to binding energies.  The requirement
of equal integral values and an assumption of $r_{0{\rm
w}}=2.5$~fm~\footnote{This value is an example which leads
to the binding energy of deuteron when another parameter is fixed to be
$v_{0{\rm w}}=-25.5$~MeV.} leads to
\begin{equation}
 v_{0{\rm w}}=v_0 \frac{3\sqrt[]{\mathstrut
  \pi}}{4}\left(\frac{r_0}{r_{0{\rm w}}}\right)^3=-20.06~{\rm MeV}.
\label{eq7}
\end{equation}
The $XA$ potential is given by
\begin{equation}
 V(r)_{\rm w}=2\pi v_{0{\rm w}} \delta_{\rm w} \frac{1}{r} \int_0^\infty dr^\prime {r^\prime}
  \rho(r^\prime) \int _{\left|r-r^\prime\right|}^{r+r^\prime} dx~xH(x-r_{0{\rm w}}),
\label{eq8}
\end{equation}
where $H(x)$ is the Heaviside step function.

As a crude assumption, the nucleon density
$\rho(r)$ is approximately given by the undistorted one for normal nucleus.  The
folding technique to derive $XA$ potential from the $XN$ potential
[equation (\ref{eq4})] then does not exactly yield the $XA$ potential in
any mathematically-rigorous method of calculation.  In order to derive precise
results of nuclear structures or energy levels, all nucleons as well as
an $X^0$ and all interactions among them need to be taken into account
with many-body quantum mechanical calculations.  Since such calculations are
unrealistically difficult, three or four-body models for an $X^0$
particle and nuclear clusters composing the nucleus should be utilized
as were done in the case of hypernuclei~\cite{Hiyama:2002yj}.  
The assumption taken here only provides
some reasonable estimate of what an effective $X$-nucleus reaction might
look like with all the nuclear degrees of freedom frozen out.  The
folding procedure, however, might produce a more useful approximation
than in the nuclear case since the $X^0$ does not participate in the
Pauli principle among nucleons.

The nucleon density of nuclei with mass number $A \geq 2$ is
assumed to be Gaussian, i.e,
\begin{equation}
 \rho(r)=\rho(0) \exp\left[-(r/b)^2\right],
\label{eq9}
\end{equation}
where
$\rho(0)=A\pi^{-3/2} b^{-3}$ is the nucleon density at $r=0$ and
satisfies the normalization $\int \rho(r) d\bfr=A$, with $A$ the mass
number.  The parameter for the width of density, i.e., $b$, is related to the root
mean square (RMS) nuclear matter radius which should be determined from
experiments, i.e., $b=\sqrt[]{\mathstrut 2/3}r_{\rm m}^{\rm RMS}$.

The $XA$ potential in the case of the Gaussian $XN$ potential,
i.e., equation (\ref{eq5}), is then simply written as
\begin{equation}
 V(r)=\frac{v_0 \delta A r_0^3}{(r_0^2+b^2)^{3/2}} \exp\left(-\frac{r^2}{r_0^2+b^2}\right).
\label{eq10}
\end{equation}

\subsection{Nuclear Binding Energies}\label{sec21}

The BBN catalyzed by the $X$ particle is significantly sensitive to binding
energies of nuclei to an $X^0$ particle ($X$-nuclei). 
The binding to $X$ particles changes the relative energies of initial and
final states, and may even change the sign of the $Q$-value.~\cite{Kusakabe:2009jt}.  Binding energies and
eigenstate wave functions of $X$-nuclei are computed taking into account
the nuclear interaction only.  The Coulomb interaction between nuclei
and the $X^0$ particle is not included since we assume that the $X^0$
has a zero charge.  The potential is supposed to be spherically
symmetric.  We solve the two-body Shr\"{o}dinger
equation by a variational calculation using the Gaussian expansion
method~\cite{Hiyama:2003cu}, and obtain binding energies.

The two-body
Shr\"{o}dinger equation for a spherically-symmetric system is
\begin{equation}
\left(-\frac{\hbar^2}{2\mu}\nabla^2 + V(r) -E\right) \psi(r)=0~,
\label{eq11}
\end{equation}
where $\hbar$ is Planck's constant, $\mu$ is the reduced mass, $V(r)$ is
the central potential at $r$, $E$ is the energy, and $\psi(r)$ is the wave
function at $r$.  If the mass of the $X^0$ particle, i.e., $m_X$,
is much heavier
than the light nuclides, $\mu$ is approximately given by the mass of the
nuclide.

The adopted RMS nuclear matter radii and their
references are listed in columns 2 and 3 in Table \ref{tab1}.  Binding
energies of ground state $X$-nuclei are
calculated with the interaction strength $\delta$ ($\delta_{\rm w}$) and
the mass $m_X$ taken as parameters.  The obtained binding energies are used for the estimation of $Q$-values of various
reactions as described below.  Similarly, we calculate
binding energies of nuclear excited states of $^4$He$_X^\ast$ and
$^8$Be$_X^\ast$ with relative angular momentum of $L=1$ by solving
equation (\ref{eq11}) for the $L=1$ states.

\begin{table}[!t]
\caption{\label{tab1} Binding energies of $X$ particles to Nuclei}
\begin{ruledtabular}
\begin{tabular}{c|cc|cc}
 & & &\multicolumn{2}{c}{$E_{\rm Bind}$~(MeV)}\\ 
nuclide & $r_{\rm m}^{\rm RMS}$~(fm)\footnotemark[1] & Ref. & $\delta=0.1$ &
 $\delta=0.2$ \\\hline
$^1$n$_X$  & ---                 & --- & --- & --- \\ 
$^1$H$_X$  & ---                 & --- & --- & --- \\ 
$^2$H$_X$  & 1.971 $\pm$ 0.005 & \cite{Martorell:1995zz} & --- & 0.367 \\
$^3$H$_X$  & 1.657 $\pm$ 0.097\footnotemark[2] & \cite{Amroun:1994qj} & 0.688 & 4.39 \\ 
$^3$He$_X$ & 1.775 $\pm$ 0.034\footnotemark[2] & \cite{Amroun:1994qj} & 0.569 & 3.85 \\ 
$^4$He$_X$ & 1.59  $\pm$ 0.04  & \cite{Tanihata:1988ub} & 2.73 & 9.81 \\ 
$^5$He$_X$ & 2.52  $\pm$ 0.03\footnotemark[3]  & \cite{Tanihata:1988ub} & 1.95 & 6.39 \\ 
$^6$He$_X$ & 2.52  $\pm$ 0.03  & \cite{Tanihata:1988ub} & 3.14 & 9.02 \\ 
$^5$Li$_X$ & 2.35  $\pm$ 0.03\footnotemark[4]  & \cite{Tanihata:1988ub} & 2.33 & 7.48 \\ 
$^6$Li$_X$ & 2.35  $\pm$ 0.03  & \cite{Tanihata:1988ub} & 3.70 & 10.5 \\ 
$^7$Li$_X$ & 2.35  $\pm$ 0.03  & \cite{Tanihata:1988ub} & 5.14 & 13.6 \\ 
$^6$Be$_X$ & 2.33  $\pm$ 0.02\footnotemark[5]  & \cite{Tanihata:1988ub} & 3.77 & 10.7 \\ 
$^7$Be$_X$ & 2.33  $\pm$ 0.02  & \cite{Tanihata:1988ub} & 5.23 & 13.8 \\ 
$^8$Be$_X$ & 2.33  $\pm$ 0.02\footnotemark[5]  & \cite{Tanihata:1988ub} & 6.74 & 17.0 \\ 
$^9$Be$_X$ & 2.38  $\pm$ 0.01  & \cite{Tanihata:1988ub} & 7.92 & 19.4 \\ 
$^9$B$_X$  & 2.45  $\pm$ 0.10\footnotemark[6]  & \cite{Fukuda1999} & 7.45 & 18.2 \\ 
$^4$He$_X^\ast$ & 1.59  $\pm$ 0.04  & \cite{Tanihata:1988ub} & --- & 2.28 \\ 
$^8$Be$_X^\ast$ & 2.33  $\pm$ 0.02\footnotemark[5]  & \cite{Tanihata:1988ub} & 3.02 & 11.1 \\ 
\end{tabular}
\footnotetext[1]{Root mean square (RMS) nuclear matter radius.}
\footnotetext[2]{Derived by $(r_{\rm m}^{\rm RMS})^2= (r_{\rm c}^{\rm RMS})^2
  -(a_p^{\rm RMS})^2$ with $a_p^{\rm RMS}=0.875 \pm 0.007$~fm using a RMS
  proton matter radius determined in experiment as a RMS charge radius.}
\footnotetext[3]{Taken from $^6$He radius.}
\footnotetext[4]{Taken from $^6$Li radius.}
\footnotetext[5]{Taken from $^7$Be radius.}
\footnotetext[6]{Taken from $^8$B radius.}
\end{ruledtabular}
\end{table}

\begin{figure}[t]
\begin{center}
\includegraphics[width=12.0cm,clip]{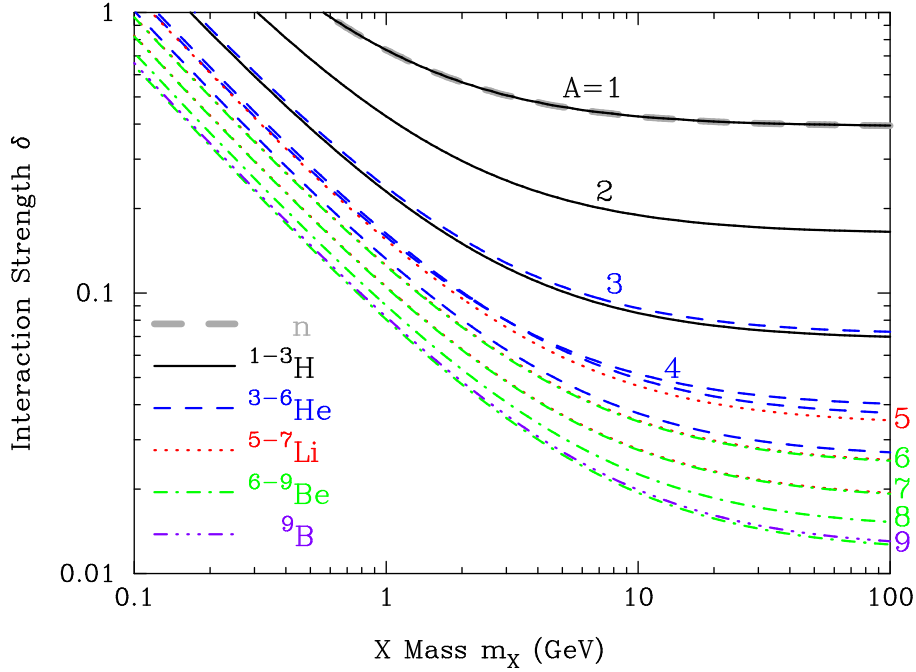}
\caption{Contours of binding energies between nuclei and an $X^0$
 corresponding to 0.1~MeV for the case of the Gaussian
 $XN$ potential.  Numbers attached to the contours indicate mass
 numbers of nuclei.\label{fig1}}
\end{center}
\end{figure}

Figure~\ref{fig1} shows the contours of binding energies of ground state
$X$-nuclei
from mass number $A=1$ to 9 in the case of the Gaussian type $XN$
potential.  The contours correspond to binding energy (BE) of
BE~$=0.1$~MeV.  This value of energy is chosen since weakly bound $X$-nuclei of
BE~$\lesssim O$(0.1~MeV) tend to be photodisintegrated by background
radiations during BBN epoch.  Such weakly bound nuclei can not attain
their large abundances without suffering from destruction processes.  In
a parameter region located at right upper side from a contour, the
$X$-nucleus can form during BBN epoch, and thus possibly affects BBN.

\begin{figure}[t]
\begin{center}
\includegraphics[width=12.0cm,clip]{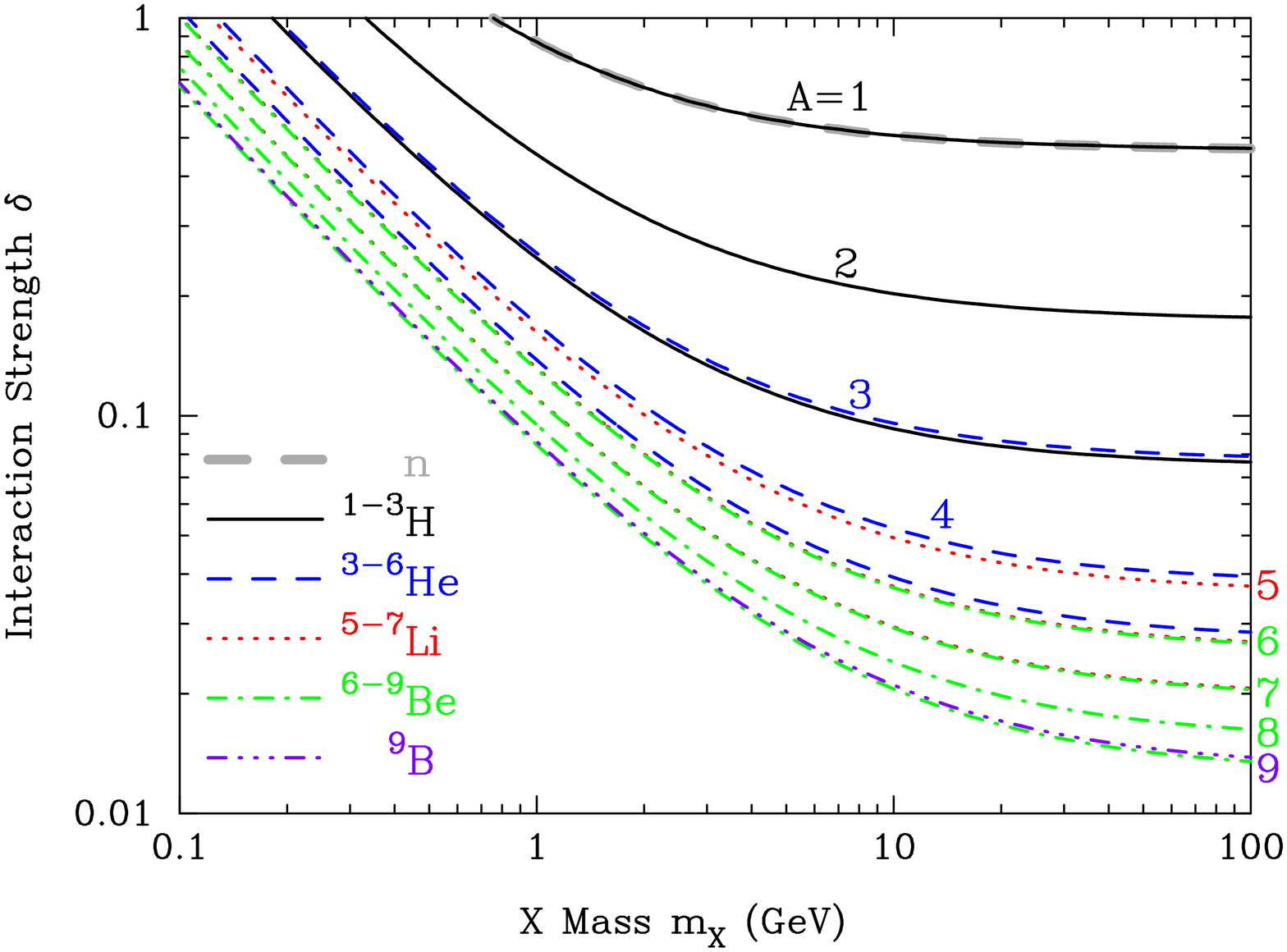}
\caption{Contours of binding energies between nuclei and an $X^0$
 corresponding to 0.1~MeV for the case of the square well
 $XN$ potential.  Numbers attached to the contours indicate mass
 numbers of nuclei.\label{fig2}}
\end{center}
\end{figure}

Figure~\ref{fig2} shows similar contours of binding energies of
BE$=0.1$~MeV in the case of the well type $XN$
potential.  The shapes of contours in both potential cases are very
similar and change slightly.  


We adopt the Gaussian $XN$ potential in calculating reaction rates and
performing a network calculation of $X$-catalyzed BBN.  After we
introduce a mechanism of $^7$Be destruction (this section) and show a
result of the nucleosynthesis for the Gaussian $XN$ potential
(Secs.~\ref{sec31} and~\ref{sec32}), we delineate parameter regions
for the $^7$Be destruction of not only the Gaussian but also the well
$XN$ potentials in Sec.~\ref{sec33}.

\subsection{Reaction Rates}\label{sec22}

Thermonuclear reaction rates $\langle \sigma v \rangle$ are roughly written (e.g., \cite{Angulo:1999zz,Boyd2008}) as
\begin{equation}
\langle \sigma v \rangle_{\rm NR} = \frac{(8/\pi)^{1/2}}{\mu^{1/2}(k_BT)^{3/2}} \int_0^\infty E \sigma(E) \exp(-E/k_BT)~dE,
\label{eq12}
\end{equation}
where
$\sigma$ is the cross section,
$v$ is the relative velocity,
$\mu$ is the reduced mass of the system,
$k_B$ is the Boltzmann constant,
$T$ is the temperature,
and
$E$ is the kinetic energy in the center of mass system.
Since there is a relation, i.e., $E=\mu v^2/2$, the equation is identical to
\begin{equation}
\langle \sigma v \rangle_{\rm NR} = \frac{2/\pi^{1/2}}{(k_BT)^{3/2}} \int_0^\infty \sigma(E)v \exp(-E/k_BT)E^{1/2}~dE.
\label{eq13}
\end{equation}
Reactions of a neutral particle and charged nuclei occur without the
effect of Coulomb repulsion.  If the quantity, i.e., $\sigma(E) v$, is
approximately given by $a+bE$ as a linear function of $E$, then the
reaction rate is simply given by
\begin{equation}
N_A \langle \sigma v \rangle_{\rm NR} = N_A \left( a+\frac{3}{2}b k_B T \right)
\label{eq14}
\end{equation}
where
the Avogadro's number $N_A=6.022\times 10^{23}$ was multiplied to both
sides of the equation.
When the product, $\sigma(E) v$, does not change drastically around
some fixed point of $E$, the integral in equation (\ref{eq13})
receives a contribution from an energy region below $E\sim k_B T$.
Information from cross sections in higher energies is, therefore, not involved
in the integral.

Reactions triggered by two charged
particles are, on the other hand affected by the Coulomb force.
The astrophysical $S$-factor is defined as
\begin{equation}
S(E)=\sigma(E)E \exp\left(\frac{2\pi Z_1 Z_2 e^2}{\hbar v}\right),
\label{eq15}
\end{equation}
where
$Z_1$ and $Z_2$ are the charge numbers of interacting particles.  The
$S$-factor might be well described by a linear function of
$E$, i.e., $S(E)=S(E_0)+\alpha(E-E_0)=S(0)+\alpha E$, where $E_0 = 0.122
(Z_1 Z_2)^{2/3} A^{1/3} T_9^{2/3}$ MeV is the most effective energy in the
integral in equation (\ref{eq12}).  We defined $A\equiv\mu/($1~amu$)$ and $T_9\equiv T/(10^9~{\rm K})$.   The reaction rate can then be written~\cite{Kamimura:2008fx} in the form of
\begin{eqnarray}
N_A \langle \sigma v \rangle_{\rm NR} &=& 7.82\times 10^6
 \left(\frac{Z_1 Z_2}{\mu}\right)^{1/3} \left(\frac{S(0)}{\rm keV b}\right) T_9^{-2/3} \exp \left[-\frac{4.25(Z_1^2 Z_2^2
				       A)^{1/3}}{{T_9}^{1/3}}\right]\nonumber \\
&&\times \left\{1+\frac{(\alpha/{\rm b})}{[S(0)/({\rm keV b})]}\left[122(Z_1^2 Z_2^2
					       A)^{1/3}
					       T_9^{2/3}+71.8T_9
					      \right]\right\}~~~~{\rm cm}^3~{\rm s}^{-1} {\rm mol}^{-1},
\label{eq16}
\end{eqnarray}
where
the Avogadro's number $N_A$ was multiplied, and
1b=10$^{-24}$ cm$^2$ was used.

We estimate rates of several important reactions in this study.  For both of
reactions by a neutral plus a charged particles and those by two charged
particles, calculated  rates are used to derive linear fitting functions
to be adopted at the energy range relevant to BBN, i.e., $T_9\lesssim
1$.

We here assume that the mass of the $X^0$ particle, i.e., $m_X$ is
100~GeV.  Calculations of nucleosynthesis are performed assuming the
Gaussian $XN$ nuclear potential as set
up in Sec.~\ref{sec21}.  We show results of the BBN
catalyzed by the $X^0$ particle for two cases of different strengths of
$XN$ interaction, i.e., $\delta=0.1$ (Case 1) and $0.2$ (Case 2) in what
follows.

In this scenario of BBN catalyzed by the $X^0$ particle, the $^7$Be can
be destroyed at its $X$-capture (Sec.~\ref{sec22a}a).  The efficiency of
this destruction, however, depends upon the fraction of
the $X^0$ particle escaping from the capture by $^4$He
(Sec.~\ref{sec22b}a).  Since other reactions of $X$-nuclei can lead to
productions of heavy nuclei, rates for such reactions are also
estimated.  Using reaction rates estimated as described in
Secs.~\ref{sec22}1--4 in our
nuclear reaction network (Sec.~\ref{sec22e}), we perform a calculation
of the catalyzed BBN.

The adopted reaction rates $N_A\langle
\sigma v \rangle$, per second per (mole cm$^{-3}$), are shown in
Tables~\ref{tab2} and \ref{tab3}.  Reaction $Q$-values are derived taking
account of the calculated binding energies of the
$X$-nuclei for Cases 1 and 2 (columns 4 and 5 in Table~\ref{tab1}).  We use the notation,
i.e., 1(2$,$3)4 for a reaction $1+2\rightarrow 3+4$.  Reaction rates related with the $X^0$ particle are estimated as follows.

\subsubsection{Non radiative reactions}\label{sec22a}

\subsubsection*{1a.   $X$($^7$Be$,^3$He)$^4$He$_X$}
This reaction is most important in this scenario.  Its reaction rate is
rather large since no radiation is involved in the reaction.  It
operates through the non resonant process unless there are resonant
states lying near the energy level of initial scattering state.  We then adopt the non resonant component of rate~\cite{Caughlan:1987qf} for the normal nuclear reaction, i.e.,
$^6$Li($n,\alpha$)$^3$H as a rough approximation.  The non resonant
component of the rate for this reaction can be extracted most easily of all
($n,\alpha$) reactions on light nuclides.  This is because heavy nuclides have large
densities of resonance so that they tend to have many resonant
components for nuclear reactions~\cite{Sowerby1970}.  The dependence of cross section on the reduced mass $\mu$ (or $A$ in
atomic mass units), i.e., $\sigma\propto A^{-2}$ has been used to
correct for de Broglie wavelengths.  

\subsubsection*{1b.   Other $X$-capture reactions}
Similarly, reaction rates of
$X$($^6$Li$,d$)$^4$He$_X$, $X$($^7$Li$,t$)$^4$He$_X$,
$X$($^7$Be$,p$)$^6$Li$_X$ and $X$($^7$Be$,n$)$^6$Be$_X$~\footnote{The $^6$Be$_X$ nucleus produced in this
reaction pathway immediately $p$-decays into $^4$He$_X$ and two
protons.} are also taken
from that of $^6$Li($n,\alpha$)$^3$H.  They are corrected for the
reduced masses.

In Case 1, the $Q$-values of $X$($^7$Be$,p$)$^6$Li$_X$ and
$X$($^7$Be$,n$)$^6$Be$_X$ are negative.  The reactions are then neglected
as well as their inverse reactions which would never become important in
changing the abundance of $^7$Be due to small abundances of $^6$Li$_X$
and $^6$Be$_X$ as shown in Section \ref{sec31}.

\subsubsection*{1c.   Destruction of $^6$Li$_X$}
The main proton burning reaction of $^6$Li$_X$ is
$^6$Li$_X$($p,^3$He)$^4$He$_X$ in Case 2.  In Case 1, on the other hand, the $Q$-value of a
three-body breakup reaction $^6$Li$_X$($p,^3$He$X$)$^4$He is
positive.  Its cross section is then much larger than that of
$^6$Li$_X$($p,^3$He)$^4$He$_X$ due to a larger phase space in the final state.  This
situation is the same as that in a similar reaction catalyzed by a
long-lived negatively charged particle $X^-$~\cite{Kamimura:2008fx}.
The reaction rates of $^6$Li$_X$($p,^3$He$X$)$^4$He (Case 1) and
$^6$Li$_X$($p,^3$He)$^4$He$_X$ (Case 2) were both taken from that of
$^6$Li($p,\alpha$)$^3$He.  In the expression for a nonresonant contribution to the thermonuclear reaction rate
[equation (\ref{eq16})], the
reduced mass was corrected.  The survival of $^6$Li$_X$
thus differs from that of $^6$Li due to only the change in reduced masses
of initial states relative to $^6$Li($p,\alpha$)$^3$He.

\subsubsection*{1d.   Production of $^9$Be$_X$}
Since the reaction $^4$He$_X$($\alpha,\gamma$)$^8$Be$_X$ is found to be
responsible for an accumulation of $^8$Be$_X$ both in Cases 1 and 2 (see
Sec.~\ref{sec3}), flows of nuclear abundances toward higher mass numbers
should be calculated.  The most important reactions in this regard are
$^8$Be$_X$($d,p$)$^9$Be$_X$ and
$^8$Be$_X$($d,n$)$^9$B$_X$~\cite{Kusakabe:2009jt}.  The reaction rates of
$^8$Be$_X$($d,p$)$^9$Be$_X$ is taken from that of
$^7$Be$_X$($d,p\alpha$)$^4$He$_X$ corrected for the reduced
mass~\cite{Kusakabe:2009jt}.  The $Q$-values of $^8$Be$_X$($d,n$)$^9$B$_X$
are negative.  This reaction can, therefore, be neglected in
environments of relatively low temperatures such as BBN.

\subsubsection{Radiative nuclear reactions}\label{sec22b}
We estimate rates of radiative capture reactions which can be important
to leave signatures of the $X^0$ particle on primordial
abundances of light elements.  Wave functions of bound and continuum
states are derived with the code RADCAP published by
Bertulani~\cite{Bertulani:2003kr} which was modified and given proper
input parameters described below.  The code also calculates cross sections of forward and
reverse reactions, and astrophysical $S$-factors.  

\subsubsection*{2a.   $X$($\alpha,\gamma$)$^4$He$_X$}\label{sec22b1}
This reaction is very important for the existence of parameter region in
which the primordial $^7$Li abundance is reduced.  The abundance of the free $X^0$
decreases if the strongly
interacting $X^0$ particle is quickly captured by $^4$He after $^4$He nuclei are
produced abundantly in the BBN epoch.  Heavy nuclei such as $^{6,7}$Li and $^7$Be whose abundances
build up after the $^4$He production are then not affected by free $X^0$
particles.  The $^4$He$_X$ nuclei can not react efficiently with Li and
Be nuclei due to large Coulomb repulsion forces by them.

Note also
that the reaction between $^4$He and $X^0$ is mainly through a radiative
capture.  The critical binding energies of $^3$H and $^3$He to an $X^0$ at which
$Q$-values of $X$($\alpha,N$)$^3A_X$ reactions change from negative to
positive are very high (see Table~\ref{tab4}).  These high binding
energies are not realized in the cases of relatively weak $XN$
interaction considered in this investigation.

Whether a given reaction operates efficiently enough to change an
abundance of relevant particle species is roughly determined by a
comparison of its rate $\Gamma$ and the cosmic expansion rate $H$.  For
a reduction of the $X^0$ abundance via a radiative capture by $^4$He, the
$\Gamma/H$ ratio is given by
\begin{equation}
\frac{\Gamma}{H}=\left(\frac{Y}{0.25}\right) \left(\frac{\eta}{6.2\times
		  10^{-10}}\right) \left(\frac{T}{0.1~{\rm MeV}}\right)
		  \left(\frac{N_A \langle \sigma v \rangle}{1.9\times
		   10^3~{\rm cm^3}~{\rm s}^{-1}~{\rm mol}^{-1}}\right),
\label{eq17}
\end{equation}
where
$Y$ is the ratio by mass of $^4$He to total baryon, and
$\eta$ is the baryon-to-photon ratio.
If $\Gamma/H>1$, then the reaction $X$($\alpha,\gamma$)$^4$He$_X$ quickly decreases
the abundance of the free $X^0$ particle.

In Case 1 the ground state of $^4$He$_X$ exists although excited states
do not.  Due to the selection rule for electromagnetic multipole
transitions, an electric dipole (E1) transition from a relative $s$-wave scattering state into the ground $s$-wave
bound state is not allowed.  The cross section then has a predominant
contribution from an E1 transition from a $p$-wave
scattering state.

In Case 2 there is one excited state of $^4$He$_X$ with relative
angular momentum of $L=1$ (see Table~\ref{tab1}).  The selection rule then allows an E1
transition from a relative $s$-wave scattering
state into the excited $p$-wave bound state.

For both cases the nuclear potential is
given by equation (\ref{eq10}) in calculating wave functions of the
ground and excited states
and scattering states of the $^4$He and $X^0$ system with the code RADCAP.  Adopting
$r_{\rm m}^{\rm RMS}=1.59$~fm, i.e., $b=1.30$~fm, the potential is given numerically
by
\begin{equation}
V_{\rm N}(r)=-123~{\rm MeV} \delta \exp\left\{-[r/(1.97~{\rm fm})]^2\right\}.
\label{eq18}
\end{equation}

\subsubsection*{2b.   $^4$He$_X$($d,\gamma$)$^6$Li$_X$}\label{sec22b2}
We calculate the radiative capture cross section with the two-body
model for the system of a $^4$He$_X$ and a deuteron.  The cross section
is proportional to the electromagnetic matrix element
squared~\cite{Bertulani:2003kr}.  The matrix element is an integral over
space of the scattering and bound state wave functions and the
operator.  The E1 operator is estimated from the initial
state wave function.  The operator is $e_1 r Y_{1\mu}(\hat{r})$ for the
radius $\bfr$ from the center of mass with the effective charge
$e_1=e(m_{^4{\rm He}_X}-2m_d)/(m_{^4{\rm He}_X}+m_d)$, and
$e_1\rightarrow e$ in the limit of an infinitely massive $X^0$ particle.  Although the wave function of the $^4$He$_X$--$d$ system
should be used in calculating the matrix element, as one
method we approximately take the wave
function of the $^6$Li--$X$ bound state calculated with equation
(\ref{eq19}) as that of the $^4$He$_X$--$d$ bound state.  As described hereinbelow, we try another method in which the wave
function of the $^6$Li$_X$ nucleus is calculated by the two-body model
for the $^4$He$_X$--$d$ system with its interaction tuned to
reproduce binding energies of $^6$Li$_X$.  Three-body
($^4$He, $d$ and $X^0$) quantum mechanical calculations are necessary
for more consistent estimations of cross sections without the approximation.

The nuclear potential for the ground state of $^6$Li$_X$ in
the final state is
given by equation (\ref{eq10}) in calculating the bound state wave
function only.  Adopting
$r_{\rm m}^{\rm RMS}=2.35$~fm, i.e., $b=1.92$~fm, the potential is given numerically
by
\begin{equation}
V_{\rm N}(r)=-99.2~{\rm MeV} \delta \exp\left\{-[r/(2.43~{\rm fm})]^2\right\}.
\label{eq19}
\end{equation}

The nuclear potential for initial scattering
states of $^4$He$_X$ and $d$ was taken from that between $X^0$
and $d$ and that between $^4$He which is bound to the $X^0$ and $d$.  The former is given by equation
(\ref{eq10}).  Adopting
$r_{\rm m}^{\rm RMS}=1.97$~fm, i.e., $b=1.61$~fm, the potential is given
numerically by
\begin{equation}
V_{\rm N}^{X-d}(r)=-45.0~{\rm MeV} \delta \exp\left\{-[r/(2.19~{\rm fm})]^2\right\}.
\label{eq101}
\end{equation}
The latter is given by
\begin{eqnarray}
 V(r)_{\rm N}^{\alpha({\rm B})-d}&=&\int \rho(\bfr^\prime)
  V_{\rm N}^{\alpha-d}(\bfx)~d\bfr^\prime \nonumber\\
&=&\frac{2\pi}{r} \int_0^\infty dr^\prime~r^\prime
 \rho(r^\prime)
 \int_{\left|r-r^\prime\right|}^{r+r^\prime} dx~x V_{\rm N}^{\alpha-d}(x).
\label{eq102}
\end{eqnarray}
where
$\bfr$ is the radius from the center of mass of $^4$He$_X$
to that of $d$,
$\bfr^\prime$ is the distance between the position of $^4$He$_X$ to that
of the $^4$He nucleus,
$\bfx=\bfr+\bfr^\prime$ is the distance between the deuteron and $^4$He,
and $\rho(\bfr^\prime)$ is the distribution function of $^4$He,
$V_{\rm N}^{\alpha-d}(x)$ is the nuclear potential between a free
$\alpha$ particle and a deuteron,
and spherical symmetries in $\rho(\bfr^\prime)$ and $V_{\rm
N}^{\alpha-d}(\bfx)$ were assumed in the last equality.

The potential $V_{\rm N}^{\alpha-d}$ is given by the two-component Gaussian function
\begin{equation}
V_{\rm N}^{\alpha-d}(r)=\sum_{i=1}^2 V_i\exp\left[-(r/r_i)^2\right],
\label{eq20}
\end{equation}
where $V_1=500$~MeV, $r_1=0.9$~rm, $V_2=-64.06$~MeV, and
$r_2=2.0$~fm~\cite{Kamimura:2008fx}.  The potential $V(r)_{\rm N}^{\alpha({\rm
B})-d}$ is then rewritten in the form of
\begin{equation}
 V(r)_{\rm N}^{\alpha({\rm B})-d}=\frac{\pi}{r} \sum_i V_i r_i^2 \int_0^\infty dr^\prime~r^\prime \rho(r^\prime)
 \left\{\exp\left[-\left(\frac{r-r^\prime}{r_i}\right)^2\right]-\exp\left[-\left(\frac{r+r^\prime}{r_i}\right)^2\right]\right\}.
\label{eq103}
\end{equation}

The Coulomb potentials for initial scattering
states of $^4$He$_X$ and $d$ originate only from that between $^4$He (inside
$^4$He$_X$) and $d$.  It is approximately given~\cite{Kamimura:2008fx} by
\begin{equation}
 V_{\rm C}^{\alpha({\rm B})-d}(r)=Z_{^4{\rm He}_X} Z_d e^2
  \frac{\mathrm{erf}(r/\sqrt[]{\mathstrut b_{^4{\rm He}_X}^2+ b_d^2})}{r},
\label{eq104}
\end{equation}
where
$Z_{^4{\rm He}_X}=2$ is the electric charge of $^4$He$_X$ nucleus,
$b_{^4{\rm He}_X}$ and $b_d$ are ranges for charges of $^4$He$_X$ and
deuteron, respectively.  $b_d=1.47$ fm is assumed which is derived from
$b_d=\sqrt[]{\mathstrut 2/3} r_{\rm C}^{\rm RMS}$ and the RMS charge radius
value of $r_{\rm C}^{\rm RMS}=1.4696$ fm~\cite{Simon:1981br}.  We take
as the $b_{^4{\rm He}_X}$ value the radius at which the charge density
$\rho_{\rm C}(r)$ of $^4$He$_X$ is $\exp(-2)$ times the maximum value,
i.e, $\rho_{\rm C}(0)$.  

The charge density of $A_X$ is given by
\begin{eqnarray}
\rho_{{\rm C},{A_X}}(r)&=& \int \rho_{A_X}(\bfr^\prime) \rho_{{\rm
 C},A}(\bfr^{\prime\prime}) d\bfr^\prime \nonumber \\
&=&\frac{2Ze}{\sqrt[]{\mathstrut \pi}br} \int_0^\infty dr^\prime~r^\prime
 \rho_{A_X}(r^\prime)
 \left\{\exp\left[-\left(\frac{r-r^\prime}{b}\right)^2\right]-\exp\left[-\left(\frac{r+r^\prime}{b}\right)^2\right]\right\},
\label{eq105}
\end{eqnarray}
where
$\bfr^\prime$ is the radius from the center of mass of
$A_X$ to that of $A$,
$\bfr^{\prime\prime}$ is the position vector of charge contributed by
the nucleus,
$r=\left|\bfr^\prime+\bfr^{\prime\prime}\right|$ is the distance from the
center of mass of $A_X$,
$\rho_{A_X}(\bfr^\prime)$ is the distribution function of $A$,
and
$\rho_{{\rm C},A}(\bfr^{\prime\prime})$ is the charge density of the
nucleus.  The nuclear charge density was assumed to be $\rho_{{\rm
C},A}(\bfr^{\prime\prime})=Ze(\pi b^2)^{-3/2}
\exp\left[-(r^{\prime\prime}/b)^2\right]$ in the last equality.
The range parameter, i.e., $b$ for $^4$He is given by
$b_{^4{\rm He}}=\sqrt[]{\mathstrut 2/3} r_{\rm C}^{\rm RMS}$ and the
RMS charge radius value of $r_{\rm C}^{\rm RMS}=1.80$
fm~\cite{Tanihata:1988ub}.

The charge densities of $^4$He$_X$ are calculated, and the ranges
of charge distribution are derived: $b_{^4{\rm
He}_X}=2.97$ fm ($\delta=0.1$) and 2.64 fm ($\delta=0.2$).  They are
used in equation (24). 

\begin{figure}[t]
\begin{center}
\includegraphics[width=12.0cm,clip]{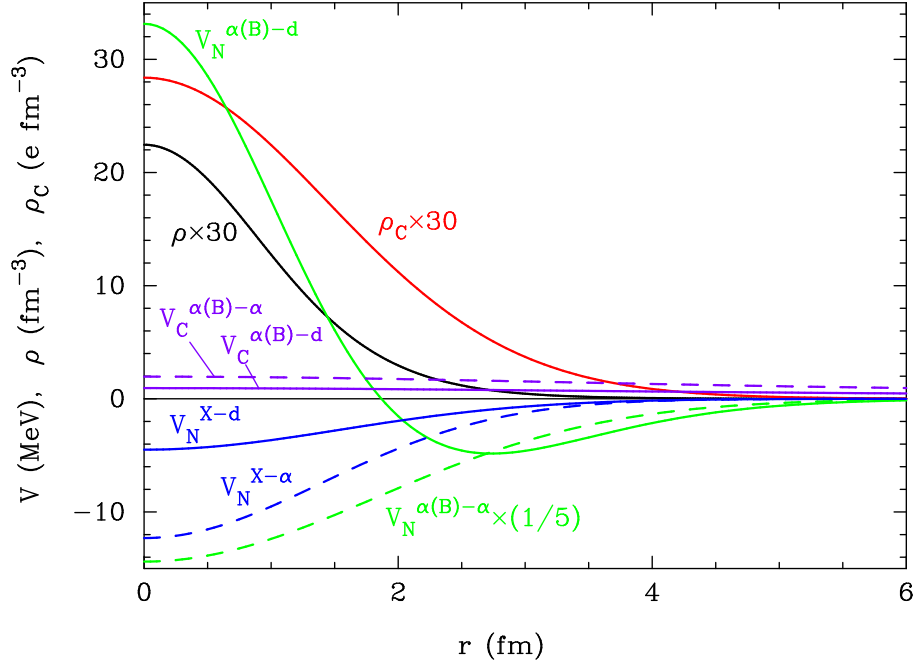}
\caption{Nuclear potentials between an $\alpha$ and a deuteron ($V_{\rm
 N}^{\alpha({\rm B})-d}$), an $X^0$ and a deuteron ($V_{\rm
 N}^{X-d}$), and the Coulomb potential between an $\alpha$ and a deuteron
 ($V_{\rm  C}^{\alpha({\rm B})-d}$) as a function of distance from the
 center of mass of the $^4$He$_X$+$d$ system (solid lines).  The similar
 potentials for the $^4$He$_X$+$\alpha$ system (dashed lines), the
 distribution function ($\rho$), and the charge density ($\rho_{\rm C}$) of $^4$He$_X$
 (solid lines) are also drawn.  It was assumed that the mass of $X^0$ particle
 is $m_X=100$~GeV, and that the interaction strength of $XN$
 force is 0.1 times that of $NN$ force ($\delta=0.1$).\label{fig8}}
\end{center}
\end{figure}

Figure~\ref{fig8} shows the adopted nuclear potential $V_{\rm N}^{X-d}$
and the calculated distribution function, i.e., $\rho$,
and charge
density, i.e., $\rho_{\rm C}$, of $^4$He$_X$, and nuclear ($V_{\rm
 N}^{\alpha({\rm B})-d}$) and Coulomb ($V_{\rm  C}^{\alpha({\rm B})-d}$)
 potentials (solid lines) for Case 1 ($\delta=0.1$) as a function of radius from the center of
 mass of $^4$He$_X$.  The center of mass is approximated to be the
 position of $X^0$ since the mass of $X^0$ is much larger than that of
 nucleon.

\begin{figure}[t]
\begin{center}
\includegraphics[width=12.0cm,clip]{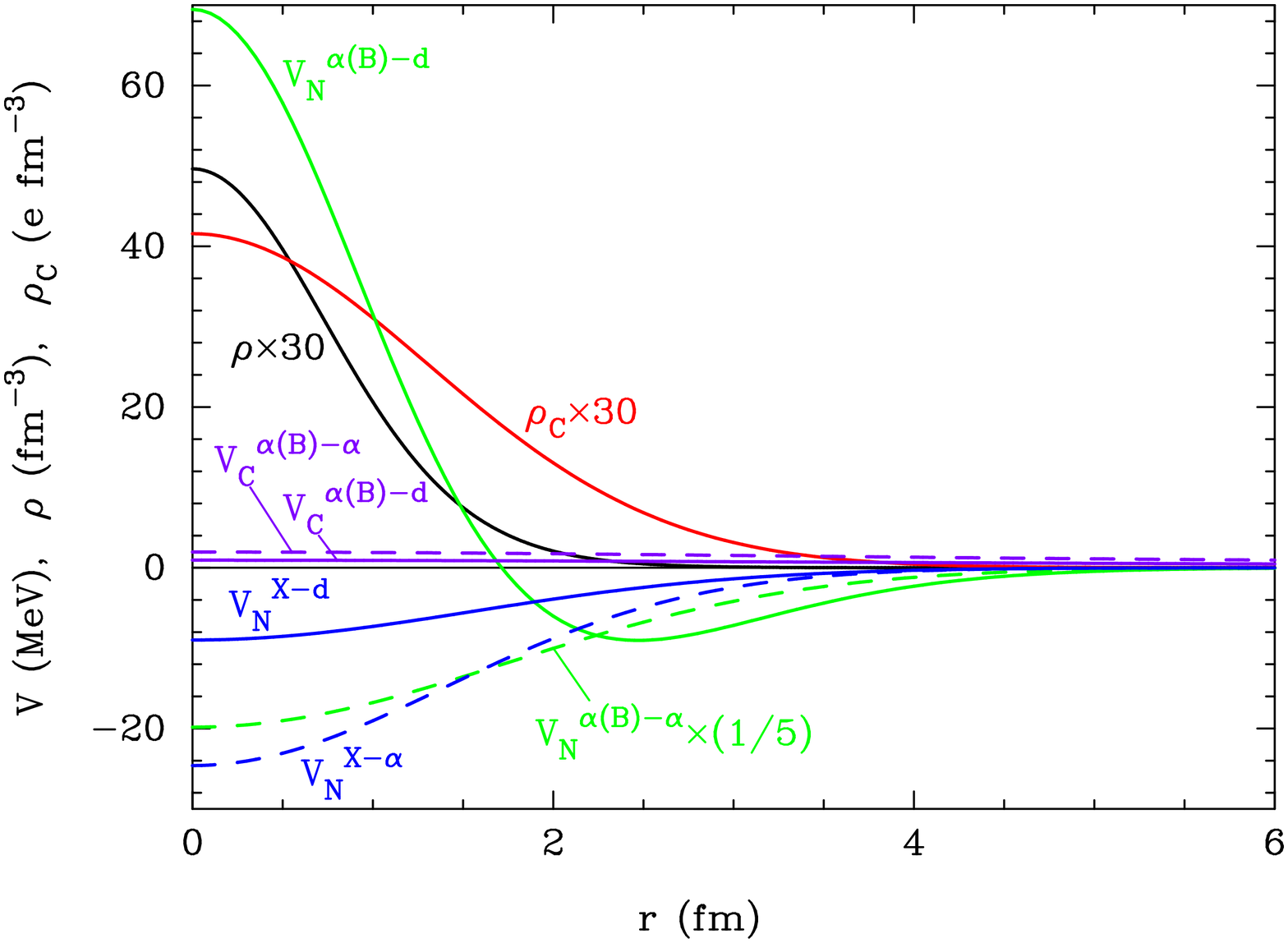}
\caption{Same as in Figure~\ref{fig8} when the interaction strength of $XN$
 force is 0.2 times that of $NN$ force ($\delta=0.2$).\label{fig9}}
\end{center}
\end{figure}

Figure~\ref{fig9} shows similar potentials and densities (solid lines)
for Case 2 ($\delta=0.2$). 

We calculate rates using another method for the same reaction.  We assume
that the $^6$Li$_X$ nucleus in the final state is described as the
two-body bound state of the $^4$He$_X$--$d$ system.  Both of initial and
final state wave functions are generated from a $^4$He--$d$ potential
tuned to reproduce the binding energy of $^6$Li$_X$ relative to separated
$^4$He$_X$ and deuteron consistent with results in Table~\ref{tab1}
[2.44 MeV for Case 1 ($\delta=0.1$) and 2.16 MeV for Case 2
($\delta=0.2$)].  We use potential terms given by equations
(\ref{eq101}), (\ref{eq103}) and (\ref{eq104}).  The parameter $\delta$ in
the nuclear potential between an $X^0$ and a deuteron [equation
(\ref{eq101})] is fitted:  $\delta_{\rm fit}=0.494$ for Case 1 and
$\delta_{\rm fit}=0.373$ for Case 2.  Calculated reaction rates are as
follows:  
\begin{equation}
N_A \langle \sigma v \rangle = 1.9\times 10^{4} T_9^{-2/3}\exp(-8.45/T_9^{1/3})
 (1+2.6 T_9^{2/3}+0.78 T_9)~~~~{\rm cm}^3~{\rm s}^{-1} {\rm mol}^{-1}
\label{eq_add1}
\end{equation}
for Case 1 ($\delta=0.1$), and
\begin{equation}
N_A \langle \sigma v \rangle = 1.7\times 10^{4} T_9^{-2/3}\exp(-8.45/T_9^{1/3})
 (1+3.6 T_9^{2/3}+1.1 T_9)~~~~{\rm cm}^3~{\rm s}^{-1} {\rm mol}^{-1}
\label{eq_add2}
\end{equation}
for Case 2 ($\delta=0.2$).  These rates differ from our standard rates
in Tables~\ref{tab2} and~\ref{tab3} by 17--26 \% (Case 1) and -45-- -33
\% (Case 2), respectively, in the important temperature range of
$T_9=$0.5--1 corresponding to $E_0=$0.15--0.25~MeV.

\subsubsection*{2c.   $^4$He$_X$($\alpha,\gamma$)$^8$Be$_X$}\label{sec22b3}
In both of Cases 1 and 2 there is one excited state of $^8$Be$_X$ with
$L=1$ below the energy level of the initial separation channel of $^4$He$_X$ and $\alpha$.  The
excitation energies are 3.72 MeV ($\delta=0.1$) and 5.89 MeV
($\delta=0.2$), respectively.  Both rates
for final states of the $^8$Be$_X$ ground and excited states are then
calculated.

The nuclear potential for the ground state of $^8$Be$_X$ in
the final state is
given by equation (\ref{eq10}) in calculating the wave function.  Adopting
$r_{\rm m}^{\rm RMS}=2.33$~{\rm fm}, i.e., $b=1.90$~fm, the potential is given numerically
by
\begin{equation}
V_{\rm N}(r)=-134~{\rm MeV} \delta \exp\left\{-[r/(2.41~{\rm fm})]^2\right\}.
\label{eq21}
\end{equation}

The nuclear potential for initial scattering states of $^4$He$_X$ and $\alpha$ was
approximately taken from that between two $\alpha$ particles given by
the three-component Gaussian function
\begin{equation}
V_{\rm N}(r)=\sum_{i=1}^3 V_i\exp\left[-(r/r_i)^2\right],
\label{eq22}
\end{equation}
where $V_1=-1.742$~MeV, $r_1=3.00$~rm, $V_2=-395.9$~MeV, $r_2=1.90$~rm,
$V_3=299.4$~MeV, and $r_3=1.74$~fm~\cite{Hiyama:2002yj}. 

The adopted nuclear potential $V_{\rm N}^{X-\alpha}$ and the calculated
nuclear ($V_{\rm N}^{\alpha({\rm B})-\alpha}$) and Coulomb ($V_{\rm
C}^{\alpha({\rm B})-\alpha}$) potentials (dashed lines) for Case 1
($\delta=0.1$) are shown  as a function of radius from the center of mass of
$^4$He$_X$ in Figure~\ref{fig8}.  Similar potentials (dashed
lines) for Case 2 ($\delta=0.2$) are shown in Figure~\ref{fig9}. 

We calculate cross sections by the two-body model in the same procedure
as performed in Sec.~\ref{sec22}b2.  First and second terms of reaction rates in Tables~\ref{tab2} and
\ref{tab3} are calculated rates for transitions to the ground and excited states, respectively.  The reaction
leading to the ground state is via an E1 transition
from an initial $p$-wave scattering state, while that leading to the excited state is
predominantly an E1 transition from an initial $s$-wave
scattering state.

\subsubsection{$\beta$-decay}\label{sec22c}
The reaction rate of $^6$Be$_X$(,$\beta^+ \nu_e$)$^6$Li$_X$ is estimated
using that of $^6$He(,$\beta^- \bar{\nu_e}$)$^6$Li.  It was corrected
for a phase space factor related to the reaction $Q$-value.  The rate is
then given by $\Gamma_X=(\ln 2/T_{1/2})(Q_X/Q)^5$, where $T_{1/2}$ is
the half life of $^6$He, $Q_X$ and $Q$ are the
$Q$-values for the $\beta$-decay of $^6$Be$_X$ and $^6$He,
respectively.  $T_{1/2}=(806.7\pm 1.5)$~ms and $Q=3.508$~MeV are
adopted~\cite{Tilley:2002vg}.  See Tables~\ref{tab2} and \ref{tab3} for
derived $Q_X$-values.

\begin{table*}[!t]
\caption{\label{tab2} Reaction rates for ($m_X$, $\delta$)=(100~GeV, 0.1)}
\begin{ruledtabular}
\scriptsize
\begin{tabular}{ccccc}
Reaction &  Reaction Rate (cm$^3$~s$^{-1}$~mole$^{-1}$) & Energy (MeV)\footnotemark[1] & Reverse Coefficient\footnotemark[2] & $Q_9$\footnotemark[3] \\
\hline
$X$($^6$Li$,d$)$^4$He$_X$ & $3.8\times 10^{6}$ & --- & 4.919 & 14.556 \\
$^6$Li$_X$($p,^3$He$\alpha$)$X$ & $3.6\times 10^{10}T_9^{-2/3}\exp(-8.81/T_9^{1/3})$ & --- & --- & 35.421 \\
$^6$Be$_X$($,e^+\nu_e$)$^6$Li$_X$ & $1.9$ & --- & --- & 44.225 \\
$X$($^4$He$,\gamma$)$^4$He$_X$    & $2.3\times 10^{2}+2.6\times 10^{3} T_9$ & --- & 7.472 & 31.659 \\
$^4$He$_X$($d,\gamma$)$^6$Li$_X$   & $1.0\times 10^{4} T_9^{-2/3}\exp(-8.45/T_9^{1/3})
 (1+6.9 T_9^{2/3}+2.0 T_9)$ & 0.15-0.25 & 2.717 & 28.331 \\
$X$($^7$Li$,t$)$^4$He$_X$      & $2.9\times 10^{6}$ & --- & 6.748 & 3.034 \\
$X$($^7$Be$,^3$He)$^4$He$_X$  & $2.9\times 10^{6}$ & --- & 6.748 & 13.252 \\
$^4$He$_X$($\alpha,\gamma$)$^8$Be$_X$    & $7.8\times 10^{5} T_9^{-2/3}\exp(-16.79/T_9^{1/3}) (1-0.39
 T_9^{2/3}-0.058 T_9)$ & 0-1 & 7.487 & 45.509 \\
 & $+4.8\times 10^{5} T_9^{-2/3}\exp(-16.79/T_9^{1/3})
 (-1+2.1 T_9^{2/3}+0.32 T_9)$ & 0.3-0.5 & & \\
$^8$Be$_X$($d,p$)$^9$Be$_X$   & $9.8\times 10^{11}  T_9^{-2/3}\exp(-13.49/T_9^{1/3})$\footnotemark[4] & --- & 1.060 & 7.242 \\
\end{tabular}
\footnotetext[1]{Rates are estimated by linear fits of
 $S$-factors in the energy ranges.}
\footnotetext[2]{For nuclides $a=i, j, k, ... $ with mass numbers $A_a$
 and numbers of magnetic substates $g_a$, the reverse coefficients are defined
 as in Ref.~\cite{Fowler:1967ty}: $0.9867(g_i g_j/g_k)(A_{i_X} A_j/A_{k_X})^{3/2}$ for the
 process $i_X$($j$,$\gamma$)$k_X$, and $[g_i g_j/(g_k
 g_l)][A_{i_X}A_j/(A_k A_{l_X})]^{3/2}$
 for the process $i_X$($j$,$k$)$l_X$.}
\footnotetext[3]{$Q_9\equiv 11.605(Q/{\rm MeV})$.}
\footnotetext[4]{The rate in Ref.~\cite{Kusakabe:2009jt} is taken.}
\end{ruledtabular}
\end{table*}

\begin{table*}[!t]
\caption{\label{tab3} Reaction rates for ($m_X$, $\delta$)=(100~GeV, 0.2)}
\begin{ruledtabular}
\scriptsize
\begin{tabular}{ccccc}
Reaction &  Reaction Rate (cm$^3$~s$^{-1}$~mole$^{-1}$) & Energy (MeV)\footnotemark[1] & Reverse Coefficient\footnotemark[2] & $Q_9$\footnotemark[3] \\
\hline

$X$($^6$Li$,d$)$^4$He$_X$ & $3.8\times 10^{6}$ & --- & 4.919 & 96.708 \\
$X$($^7$Be$,n$)$^6$Be$_X$ & $2.9\times 10^{6}$ & --- & 34.139  & 0.130 \\
$X$($^7$Be$,p$)$^6$Li$_X$       & $2.9\times 10^{6}$ & --- & 11.380 & 56.727 \\
$^6$Li$_X$($p,^3$He)$^4$He$_X$    & $3.6\times 10^{10}T_9^{-2/3}\exp(-8.81/T_9^{1/3})$ & --- & 0.593 & 38.678 \\
$^6$Be$_X$($,e^+\nu_e$)$^6$Li$_X$   & $6.2\times 10^{-10}$ & --- & --- & 44.225 \\
$X$($^4$He$,\gamma$)$^4$He$_X$       & $5.0\times 10^{4}-2.2\times 10^{3} T_9$ & --- & 7.472 & 113.811 \\
$^4$He$_X$($d,\gamma$)$^6$Li$_X$      & $8.2\times 10^{3} T_9^{-2/3}\exp(-8.45/T_9^{1/3})
 (-1+6.6 T_9^{2/3}+2.0 T_9)$ & 0.15-0.25 & 2.717 & 25.074 \\
$X$($^7$Li$,t$)$^4$He$_X$       & $2.9\times 10^{6}$ & --- & 6.748 & 85.186 \\
$X$($^7$Be$,^3$He)$^4$He$_X$     & $2.9\times 10^{6}$ & --- & 6.748 & 95.404 \\
$^4$He$_X$($\alpha,\gamma$)$^8$Be$_X$      & $2.4\times 10^{4} T_9^{-2/3}\exp(-16.79/T_9^{1/3})
 (1+0.15 T_9^{2/3}+0.022 T_9)$ & 0-1 & 7.487 & 82.366 \\
 & $+2.7\times 10^{4} T_9^{-2/3}\exp(-16.79/T_9^{1/3}) (-1+2.7
 T_9^{2/3}+0.41 T_9)$ & 0.3-0.5 & &  \\
$^8$Be$_X$($d,p$)$^9$Be$_X$      & $9.8\times 10^{11}  T_9^{-2/3}\exp(-13.49/T_9^{1/3})$\footnotemark[4] & --- & 1.060 & 20.957 \\
\end{tabular}
\footnotetext[1]{Rates are estimated by linear fits of
 $S$-factors in the energy ranges.}
\footnotetext[2]{For nuclides $a=i, j, k, ... $ with mass numbers $A_a$
 and numbers of magnetic substates $g_a$, the reverse coefficients are defined
 as in Ref.~\cite{Fowler:1967ty}: $0.9867(g_i g_j/g_k)(A_{i_X} A_j/A_{k_X})^{3/2}$ for the
 process $i_X$($j$,$\gamma$)$k_X$, and $[g_i g_j/(g_k
 g_l)][A_{i_X}A_j/(A_k A_{l_X})]^{3/2}$
 for the process $i_X$($j$,$k$)$l_X$.}
\footnotetext[3]{$Q_9\equiv 11.605(Q/{\rm MeV})$.}
\footnotetext[4]{The rate in Ref.~\cite{Kusakabe:2009jt} is taken.}
\end{ruledtabular}
\end{table*}

\subsubsection{Unimportant pathways through $^5$He$_X$ and $^5$Li$_X$}\label{sec22d}

Reactions of $^5$He$_X$ and $^5$Li$_X$ are found to be unimportant in this
study under the assumption of our models described above.

The binding of the $X^0$ particle to $^5$He and $^5$Li can lead to
stabilizations of such bound states against neutron and proton
emissions, respectively~\cite{Kusakabe:2009jt}.  This study, however, indicates that $^5A_X$ nuclei do not play a significant role in the BBN epoch in
both Cases 1 and 2.
The reason is as follows:

A sufficient condition for that the $^5A_X$ nuclei are not
important for a production of heavy nuclei of $A>6$ is satisfied in
Cases 1 and 2:  $Q$-values of reactions $^4$He$_X$($d,n$)$^5$Li$_X$ and
$^4$He$_X$($d,p$)$^5$He$_X$ are negative.  This means that $^5$Li$_X$
and $^5$He$_X$ produced during BBN (if any via radiative captures of $p$
or $n$ by $^4$He$_X$) is quickly processed by $n$ or $p$
back into $^4$He$_X$ via $^5$Li$_X$($n,d$)$^4$He$_X$ and
$^5$He$_X$($p,d$)$^4$He$_X$.  Note that the destruction reactions are
strong since no radiation is involved in the reactions.

In passing, if the energy level of $^5$He$_X$ ($^5$Li$_X$) nucleus is
lower than that
of $^4$He$_X$+$n$ ($p$) separation channel, i.e., $Q(^4$He$_X+N\rightarrow^5A_X)\gtrsim 0$~MeV, the $X$-nucleus can be produced during BBN.  Amounts of
such $X$-nuclei are, however, very small due to the reason explained in
the previous paragraph.  In this investigation the $^5A_X$ nuclides have
not been stabilized in both cases of potential types for
parameter space studied.  The $^5A_X$ nuclei then
do not play a role.  There is, however, a large uncertainty in binding
energies of $X$-nuclei which stems from $XA$ potentials, and from
precise nucleon density of $X$-nuclei.  Such
uncertain points should be studied using specific particle models
describing potentials and dedicated quantum many-body models taking
account of interactions inside the normal nuclei such as $^5A=^4$He+$N$
during the processes.

\subsubsection{Nuclear Reaction Network}\label{sec22e}

We calculate binding energies of $X$-nuclei, and
$Q$-values of reactions involving the $X^0$ particle and $X$-nuclei.
We estimate rates of several (forward and reverse) reactions which
possibly play roles in BBN as
described above, and calculated the catalyzed BBN.  We note
that productions of double $X$-nuclei, i.e., $A_{XX}$ are not taken into
account in this study.  The ratio of the rate for
reaction between $X$-nuclei and an $X^0$ to that of cosmic expansion is
given by
\begin{equation}
\frac{\Gamma}{H}=\left(\frac{n_X/n_b}{10^{-4}}\right) \left(\frac{\eta}{6.2\times
		  10^{-10}}\right) \left(\frac{T}{0.1~{\rm MeV}}\right)
		  \left(\frac{N_A \langle \sigma v \rangle}{1.2\times
		   10^6~{\rm cm^3}~{\rm s}^{-1}~{\rm mol}^{-1}}\right).
\label{eq23}
\end{equation}
An efficient operation of a reaction should satisfy $\Gamma/H>1$, which
is possible when $N_A \langle \sigma v \rangle \gtrsim
10^6$~cm$^3$~s$^{-1}$~mol$^{-1}$.  $X$-nuclei with significant
abundances during the catalyzed BBN seen in Sec.~\ref{sec3} are
$^4$He$_X$, $^6$Li$_X$, $^8$Be$_X$ and $^9$Be$_X$.  $^4$He$_X$
can react with an $X^0$ particle only via the radiative capture
since there is no exit channel of particle break up.  Since reaction
rates of radiative capture are typically small, $^4$He$_{XX}$ production
is not efficient.  The effect of $^4$He$_{XX}$ would thus be negligible
although some fraction of $^4$He$_X$ would react with an $X^0$ and form
the $^4$He$_{XX}$.  Other $X$-nuclei, i.e., $^6$Li$_X$, $^8$Be$_X$ and
$^9$Be$_X$ can, on the other hand, react nonradiatively with an $X^0$.
Possible reactions are $^6$Li$_X$($X,d$)$^4$He$_{XX}$,
$^8$Be$_X$($X,\alpha$)$^4$He$_{XX}$, $^8$Be$_X$($X,\alpha_X$)$^4$He$_X$,
$^9$Be$_X$($X,n$)$^8$Be$_{XX}$ and $^9$Be$_X$($X,\alpha n$)$^4$He$_{XX}$
(whether their $Q$-values are positive or negative should be determined by
more sophisticated estimations of binding energies).  Since abundances
of $^6$Li$_X$, $^8$Be$_X$ and $^9$Be$_X$ are rather small in cases
studied in this paper (Sec.~\ref{sec3}), their processing would have
negligible effects on final abundances of light elements.  We note,
however, that final abundances of $^6$Li$_X$ and $^9$Be$_X$ may be
reduced through nonradiative reactions with an $X^0$ particle.

The BBN network code of Refs.~\cite{Kawano1992,Smith:1992yy} is modified and
used.  The $X^0$ particles and relevant $X$-nuclei are included as new
species.  Reactions connecting normal and $X$-nuclei and the $X^0$
particle are added to the code (see Tables~\ref{tab2} and \ref{tab3} for
their rates).  Nuclear reaction rates for the SBBN~\cite{Kawano1992,Smith:1992yy} have been replaced with new rates published in
Refs.~\cite{Descouvemont:2004cw,Cyburt:2008kw} and the adopted neutron
lifetime is $\tau_n=881.9$~s~\cite{Mathews:2004kc}.

\section{Results}\label{sec3}
\subsection{Evolution of nuclear abundances}\label{sec31}

\begin{figure}[t]
\begin{center}
\includegraphics[width=12.0cm,clip]{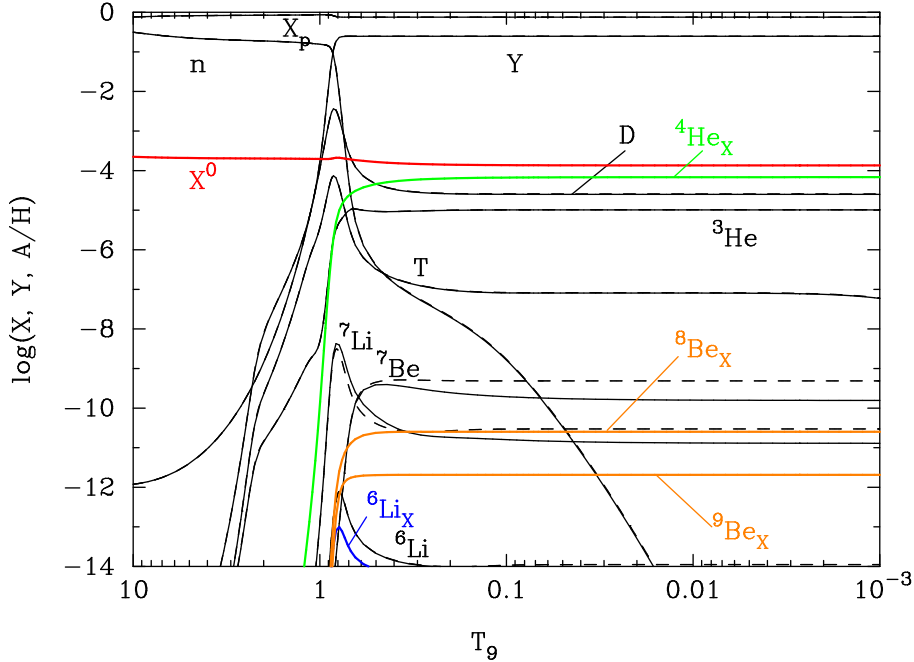}
\caption{Calculated abundances of normal and
 $X$-nuclei (solid lines) as a function of $T_9$.  The mass of $X^0$ particle
 was assumed to be $m_X=100$~GeV, and the interaction strength of $XN$
 force is 0.1 times that of $NN$ force ($\delta=0.1$).  For this figure, we took the $X^0$ abundance to be
 $N_X/n_b=1.7\times 10^{-4}$ ($Y_X\equiv N_X/s=1.5\times
10^{-14}$), and its lifetime to be much longer than BBN time scale, i.e., $\tau_X\gg 200$~s.  The $X^0$
 reaction rates are given as described in the text, Section~\ref{sec2}.  The dashed lines
 correspond to the abundances of normal nuclei in the standard BBN model.\label{fig3}}
\end{center}
\end{figure}

Figure~\ref{fig3} shows results of the abundances of normal and
$X$-nuclei as a function of temperature for Case 1 ($\delta=0.1$).  The
mass of the $X^0$ has been taken to be $m_X=100$~GeV.   Its initial
abundance is $N_X/n_b=1.7\times 10^{-4}$ ($Y_X\equiv N_X/s=1.5\times
10^{-14}$), where $N_X$ and $n_b$ are the number densities of the $X^0$
particles and baryons, respectively.  This abundance is
chosen as an example leading to a significant $^7$Be reduction.  It is
set as a parameter here since the $X^0$ abundance is very uncertain.  The decay lifetime is assumed to
be much longer than BBN time scale, i.e., $\tau_X\gg 200$~s so that
effects of the decay are not seen.  The $X^0$ particle is assumed to
have been long extinct by now.


At high temperatures $T_9 \agt 1$, the $X^0$ particles exist mainly in
the free state since efficient photodisintegrations of $X$-nuclei destroy
the bound state.  At $T_9\sim 1$ the $^4$He synthesis occurs as in
SBBN,
and about one third of $X^0$ particles are then captured by $^4$He nuclei.
$^4$He$_X$ nuclei produced in this epoch react with normal nuclei, and
affect abundances of $^7$Li [by $^4$He$_X$($t,^7$Li)$X$~\footnote{This
reaction can be efficient since the reaction $Q$-value of
$X$($^7$Li$,t$)$^4$He$_X$ is very small (see Table~\ref{tab2}).}], $^6$Li$_X$ [by
$^4$He$_X$($d,\gamma$)$^6$Li$_X$], and $^8$Be$_X$ [by
$^4$He$_X$($\alpha,\gamma$)$^8$Be$_X$].  Note that $^6$Li$_X$ nuclei produced at $T_9\sim
1$ experience a strong destruction process, i.e.,
$^6$Li($p,^3$He$\alpha$)$X$.  $^9$Be$_X$ is produced by
$^8$Be$_X$($d,p$)$^9$Be$_X$.  At last, the most important processes
operate.  Free $X^0$ particles which survived the
capture by $^4$He react with $^7$Be [by $X$($^7$Be$,^3$He)$^4$He$_X$]
and $^7$Li [by $X$($^7$Li$,t$)$^4$He$_X$].  The abundances of $^7$Be and
$^7$Li thus decrease.

\begin{figure}[t]
\begin{center}
\includegraphics[width=12.0cm,clip]{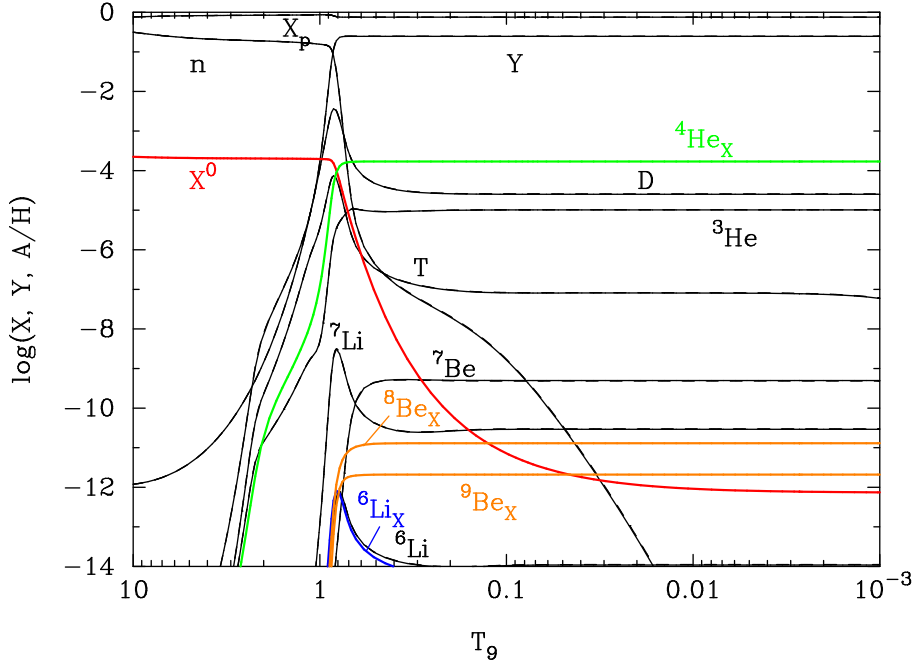}
\caption{Same as in Figure 3 when the interaction strength of $XN$
 force is 0.2 times that of $NN$ force ($\delta=0.2$).  \label{fig4}}
\end{center}
\end{figure}

Figure~\ref{fig4} shows results of the abundances of normal and
$X$-nuclei as a function of temperature for Case 2 ($\delta=0.2$).
Parameters other than $\delta$ are the same as in Figure~\ref{fig3}.  A clear
difference from Case 1 is a complete capture of the $X^0$ particle by
$^4$He (see Section~\ref{sec22b}).  Decreases in the abundances of
$^7$Be and $^7$Li are, therefore, not seen.  The productions of $^6$Li$_X$,
$^8$Be$_X$, and $^9$Be$_X$ result following the $^4$He$_X$ production
similarly to Case 1.


\subsection{Decrease in the primordial $^7$Li abundance}\label{sec32}

\begin{figure}[!t]
\begin{center}
\includegraphics[width=9.0cm,clip]{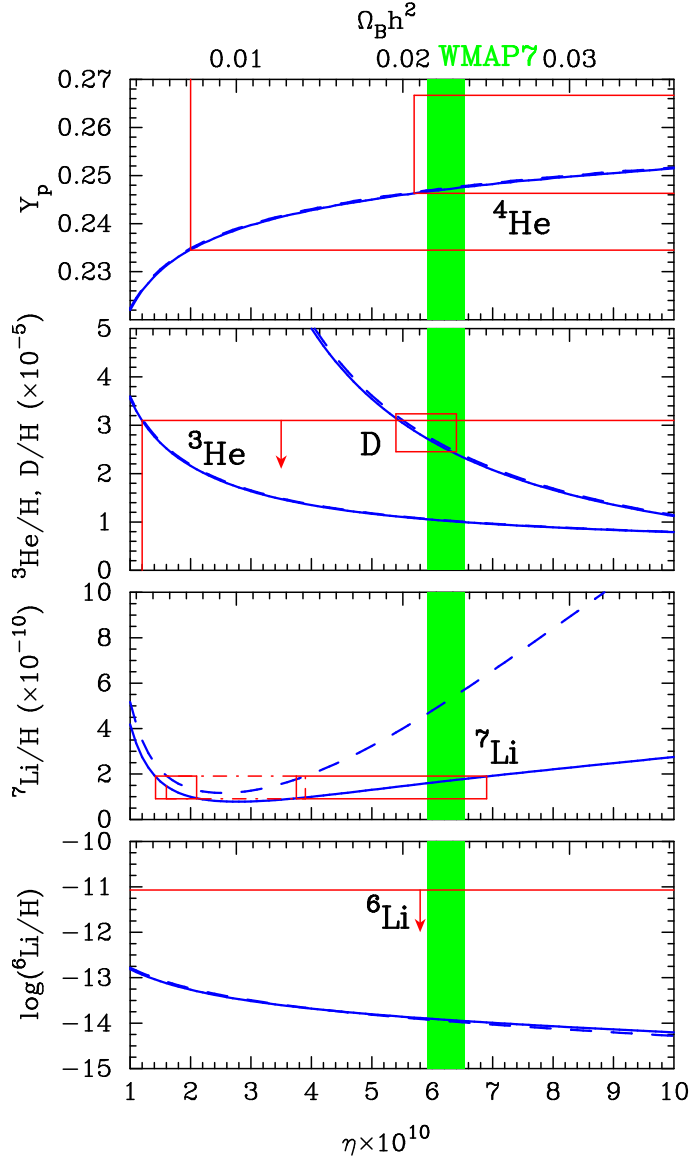}
\caption{Abundances of $^4$He (mass fraction), D, $^3$He,
 $^7$Li and $^6$Li (by number relative to H) as a function of the
 baryon-to-photon ratio $\eta$ or the baryon energy density $\Omega_B
 h^2$ of the universe.  The solid
 curves are the calculated results in the $X$ catalyzed BBN for the case of ($m_X$, $\delta$, $Y_X$,
 $\tau_X$)=(100~GeV, 0.1, $1.5\times 10^{-14}$, $\infty$), while the dashed curves
 are those in the standard BBN.  There is virtually no difference between the
 dashed and solid curves for $^4$He, D, $^3$He, and $^6$Li.  The boxes represent
 the adopted abundance constraints from Refs.~\cite{Izotov:2010ca,Aver:2010wq} for
 $^4$He,~\cite{Pettini:2008mq} for D,~\cite{Bania:2002yj} for $^3$He,~\cite{Ryan2000} for $^7$Li,
 and~\cite{Asplund:2005yt} for $^6$Li, respectively.  The vertical
 stripe represents the 2~$\sigma$~$\Omega_B h^2$ limits provided by
 WMAP~\cite{Larson:2010gs}. \label{fig5}}
\end{center}
\end{figure}

Figure~\ref{fig5} shows abundances (solid curves)
 of $^4$He (mass fraction), D, $^3$He,
 $^7$Li and $^6$Li (by number relative to H) as a function of the
 baryon-to-photon ratio $\eta$ or the baryon energy density $\Omega_B
 h^2$ of the universe.  The solid curves are the calculated result in
 the $X$ catalyzed BBN for Case 1, i.e., ($m_X$, $\delta$, $Y_X$,
 $\tau_X$)=(100~GeV, 0.1, $1.5\times 10^{-14}$, $\infty$).  The dashed curves are
 those in the SBBN.  The boxes correspond to the adopted
 constraints on primordial abundances (see Appendix~\ref{appendix}).  The vertical stripe
 shows the $2\sigma$ limits on $\Omega_{\rm
 B}h^2=0.02258^{+0.00057}_{-0.00056}$ provided by
 WMAP~\cite{Larson:2010gs} for the $\Lambda$CDM+SZ+lens model.

The decrease in the $^7$Li abundance is found, while other nuclear
abundances are not changed.  A solution to
the $^7$Li problem is thus found in this model.  One should note,
however, that the abundance of $^9$Be$_X$ could be higher than the
adopted constraint on the primordial $^9$Be abundance.  For example, in
Figure~\ref{fig3}, the final abundance of $^9$Be$_X$/H $\sim 2\times 10^{-12}$
is shown.  Effects of the decay of the $X^0$ particle inside $X$-nuclei
are not addressed in this paper.  They should be studied in order to
estimate fractions of $^9$Be$_X$ which remain as $^9$Be after the decay
of $X^0$ particle.


\subsection{Parameter region for Li reduction}\label{sec33}

If the strength of $XN$ interaction is relatively weak as in the Cases
1 and 2 which we study in this paper, most strong reactions for the $X^0$ particle
to get bound to nuclei would be non-radiative $X^0$-capture reactions
which are found important in the present model.  We,
however, note that efficiencies of the reactions are uncertain, and that
the present result is based on a rough assumption that the non-radiative
cross sections have been set equal to the $^6$Li($n,\alpha$)$^3$H cross
section excepting for the factor of reduced mass.  
Although effects of heavy $X^0$ particles during BBN was first studied
recently~\cite{Kusakabe:2009jt}, the possibility of $X^0$ capture reactions
via nucleon emission has been mentioned in 1995~\cite{Plaga1995}.  In
order for these reactions to occur efficiently during BBN, reaction
$Q$-values need to be positive.  The $Q$-value of the reaction
$X$($A,b$)$B_X$ is given by
\begin{equation}
Q={\rm BE}(B)+{\rm BE}(B_X)+{\rm BE}(b)-{\rm BE}(A),
 \label{eq24}
\end{equation}
where
BE$(A)$, BE$(B)$ and BE$(b)$ are the binding energies of
nuclei $A$, $B$ and $b$ with
respect to separated nucleons, respectively,
and BE$(B_X)$ is the binding energies of $B_X$ with
respect to separated $B$ and $X^0$.

In Table~\ref{tab4} critical binding energies of $B_X$ realizing $Q>0$ are
listed.  We have taken data on binding energies of
normal nuclei from TUNL Nuclear Data Evaluation
Project~\footnote{WWW:
http://www.tunl.duke.edu/nucldata/index.shtml.}.

\begin{table}[!t]
\caption{\label{tab4} Critical binding energies of $X$-nuclei}
\begin{ruledtabular}
\begin{tabular}{ccc}
Number & Reaction & Binding energy (MeV) \\\hline
1 & $X$($^2$H$,n$)$^1$H$_X$   & 2.224 \\
2 & $X$($^2$H$,p$)$^1n_X$     & 2.224 \\
3 & $X$($^3$H$,n$)$^2$H$_X$   & 6.257 \\
4 & $X$($^3$He$,p$)$^2$H$_X$ & 5.494 \\
5 & $X$($^4$He$,p$)$^3$H$_X$  & 19.815 \\
6 & $X$($^4$He$,n$)$^3$He$_X$ & 20.578 \\
7 & $X$($^6$Li$,p$)$^5$He$_X$ & 4.497 \\
8 & $X$($^6$Li$,n$)$^5$Li$_X$ & 5.39 \\
9 & $X$($^7$Li$,p$)$^6$He$_X$ & 9.975 \\
10 & $X$($^7$Li$,n$)$^6$Li$_X$ & 7.250 \\
11 & $X$($^7$Be$,p$)$^6$Li$_X$ & 5.606 \\
12 & $X$($^7$Be$,n$)$^6$Be$_X$ & 10.676 \\
13 & $X$($^4$He$,d$)$^2$H$_X$ & 23.848 \\
14 & $X$($^6$Li$,d$)$^4$He$_X$ & 1.474 \\
15 & $X$($^7$Li$,t$)$^4$He$_X$ & 2.467 \\
16 & $X$($^7$Be$,^3$He)$^4$He$_X$ & 1.587 \\
\end{tabular}
\end{ruledtabular}
\end{table}

More important reactions than nucleon emissions in this $X$ catalyzed
BBN scenario are $X$($^6$Li$,d$)$^4$He$_X$, $X$($^7$Li$,t$)$^4$He$_X$, and
$X$($^7$Be$,^3$He)$^4$He$_X$~\footnote{The reaction rates and related information of $^4$He$_X$($d,^6$Li)$X$,
$^4$He$_X$($t,^7$Li)$X$ and $^4$He$_X$($^3$He$,^7$Be)$X$ in the case of
leptonic $X^-$ particle can be found in Refs.~\cite{Hamaguchi:2007mp,Kamimura:2008fx}}.  Critical binding energies realizing $Q>0$
for emissions of particles other than nucleons are also listed in
Table~\ref{tab4}.

\begin{figure}[t]
\begin{center}
\includegraphics[width=12.0cm,clip]{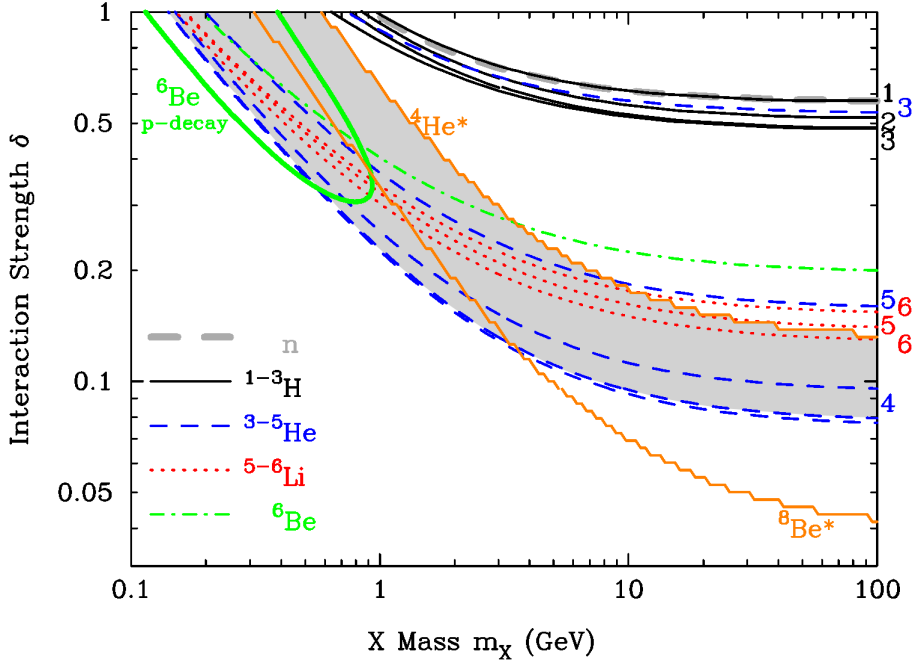}
\caption{Contours of binding energies between nuclei and an $X^0$
 particle corresponding to $Q=0$ of reactions (thin and thick smooth
 curves) for the case of the Gaussian $XN$ potential.  Numbers attached
 to the contours indicate mass  numbers of nuclei.  Above the contours, reaction $Q$-values are
 positive (see text, Sec~\ref{sec33}).  Zigzag curves correspond to boundaries
 above which an excited state of $^4$He$^\ast$ (upper line) and
 $^8$Be$^\ast$ (lower) exist, respectively.  In the shaded region, the
 $^7$Li problem can be resolved. \label{fig6}}
\end{center}
\end{figure}

Figure~\ref{fig6} shows contours in the parameter space ($m_X$,
 $\delta$) for critical binding energies of $X$-nuclei (thin and thick smooth curves) for the case of the Gaussian $XN$ potential.  Critical
 binding energies chosen for the plot are those of reactions, whose numbers
 are defined in Table~\ref{tab4}: 2 (for $n_X$), 1 ($^1$H$_X$) 3, 4 ($^2$H$_X$), 5 ($^3$H$_X$), 6
 ($^3$He$_X$), 14, 15, 16 ($^4$He$_X$), 7 ($^5$He$_X$), 8 ($^5$Li$_X$),
 10, 11 ($^6$Li$_X$), 12 ($^6$Be$_X$).  Numbers attached
 to the contours indicate mass numbers of nuclides for elements which have more than
 two isotopes plotted.  The contour of $Q=0$ for the proton decay of $^6$Be$_X$,
 i.e., $^6$Be$_X$($,2p$)$^4$He$_X$, is also shown as a thick solid
 line.  Above the contours, reaction $Q$-values are
 positive.  Zigzag curves correspond to boundaries
 above which an excited state of $^4$He$^\ast$ with $L=1$ (upper line) and
 $^8$Be$^\ast$ with $L=1$ (lower) exist, respectively.


\begin{figure}[t]
\begin{center}
\includegraphics[width=12.0cm,clip]{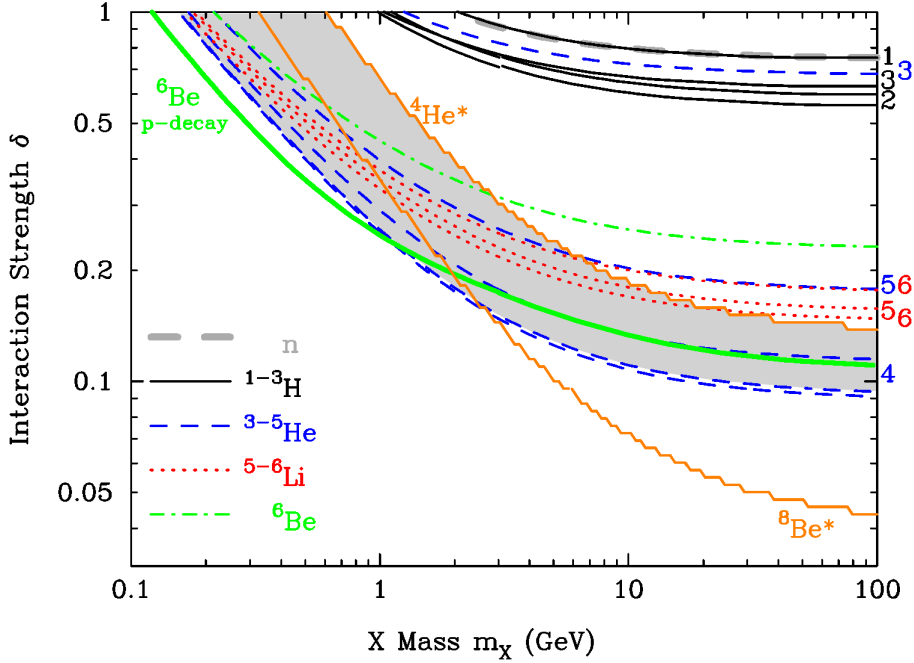}
\caption{Same as in Figure \ref{fig6} for the case of the square well
 $XN$ potential. \label{fig7}}
\end{center}
\end{figure}

Figure~\ref{fig7} shows contours for the case of the square well $XN$
 potential corresponding to the same boundaries as
 in Figure~\ref{fig6}.  The contours are very similar to those in
 Figure~\ref{fig6} excepting for that of the proton decay of
 $^6$Be$_X$.  Although the difference in the contours of $^6$Be$_X$ does
 not affect the BBN significantly, a realistic estimation of binding
 energies regarding all light nuclei are very important to obtain a
 realistic result of the catalyzed BBN.


We delineate the parameter region which might be responsible for a
reduction of the primordial $^7$Li abundance.  The
baryon-to-photon ratio is fixed to be $\eta=6.225\times 10^{-10}$ from WMAP
determination~\cite{Larson:2010gs}.  In both
Figures \ref{fig6} and \ref{fig7}, it is found that the contours of the boundaries for 
existences of $^4$He$_X^\ast$ are above the contours of the reaction
$X$($^7$Be$,^3$He)$^4$He$_X$ (the second lowest thin dashed lines).  In the
parameter region in right upper side from the curve of $^4$He$_X^\ast$,
free $X^0$ particles are captured onto $^4$He at $T_9\sim 1$ before
they can react with $^7$Be to reduce its abundance.

In the shaded region
below that curve and above the curve of $X$($^7$Be$,^3$He)$^4$He$_X$, some
amount of free $X^0$s possibly remain, and they can reduce the $^7$Be
abundance.  This shaded region is, therefore, a possible parameter region where
the $^7$Li problem is solved.  For a significant
destruction of $^7$Be, however, a relatively large value of initial
abundance, i.e., $Y_X\sim O(10^{-14})$ is needed.  Under the assumption
of thermal freeze-out from equilibrium of the $X^0$ particle, the initial
(relic) abundance is large if the annihilation cross section of $X^0$ particle
is small [c.f. equation (\ref{eq2})].  The required abundance may
then realize in the scenario of sub-strongly interacting particle $X^0$ which
has survived the annihilation due to its small interaction strength.

\section{Conclusions and discussion}\label{sec4}

We have investigated effects on BBN of a long-lived strongly interacting massive
particle (SIMP) $X^0$ for different masses $m_X$ and strengths of $XN$
interaction, i.e., $\delta$.
Binding energies of bound states of nuclei and an $X^0$ particle, i.e.,
$X$-nuclei, are calculated for two types of $XN$ potentials, i.e., Gaussian
and well types.  It is shown that calculated binding energies are not
largely dependent upon the potential shapes, and are determined by the
interaction strength at a given mass of $X^0$.

Evolutions of light element abundances are calculated as a function of
the temperature for two specific cases of relatively weak interaction
strengths.  We found that $^7$Be and $^7$Li can be
destroyed by the nuclear capture reactions of free $X^0$ particles.  The
reactions identified as destruction processes are
$X$($^7$Be$,^3$He)$^4$He$_X$ and $X$($^7$Li$,t$)$^4$He$_X$.  We show that
the lack of an excited state of $^4$He$_X$ with a relative angular
momentum $L=1$ is essential for some fraction of the $X^0$
particles to escape capture by $^4$He.

We suggest that the $^7$Li problem could be solved based upon a
net work calculation of catalyzed BBN, and found the parameter
region in the ($m_X$, $\delta$) plane where the $^7$Li problem can be fixed.

We note that the results have been derived under the
assumption that the $X^0$ particles do not change nuclear structure apart
from sticking unaltered nuclei to the particles.  This rough
approximation is unlikely to be true especially in the case of
relatively large strength of interaction since the $XN$ potential is not
much weaker than the $NN$ potential and can not be neglected.  More
realistic estimations of wave functions and binding energies of
$X$-nuclei need to include such changes in nuclear structures with
the use of three or more-body models.  In the case of strongly bound
$^4$He, the binding energy of the ground state is 28 MeV which is larger
than binding energy of $^4$He$_X$ with respect to separated $^4$He and
$X^0$.  The effect of change in nuclear density caused by the existence
of the $X^0$ then tends to be small for $^4$He.  The calculated binding
energy of $^4$He$_X$ and the cross section for radiative capture of
$X^0$ on $^4$He are, therefore, likely to be accurate.  If the $XN$ and $XA$ potentials adopted in this paper
describe well the real interaction, the main uncertainty in the BBN
calculation would be in the estimations of nuclear reactions
involving the $X^0$ particle.  In this paper, the cross
section of the most important reaction, i.e.,
$X$($^7$Be,$^3$He)$^4$He$_X$ was estimated only by analogy with
$^6$Li($n,\alpha$)$^3$H, and many other cross sections were also estimated
using standard nuclear reaction rates.  Radiative reaction cross
sections were estimated approximately within the framework of two-body
models.  The rates should, however, be calculated in more
rigorous quantum many body models, not by the rough Born approximation, in
order to derive realistic values.  Although there might be errors
in the calculated abundances in the $X$-catalyzed BBN of one order of
magnitude or so, we argue at this moment that there is a possibility of $^7$Li
reduction in the BBN model including a long-lived sub-SIMP.

Effects of possible direct interactions of decay products of $X^0$ with the remaining
nuclei $A$ at the decay of $X^0$ in an $X$-nucleus $A_X$ are not
taken into account yet.  They should be studied in the future in order to better estimate final
abundances of the light elements.  In addition,
nonthermal nucleosynthesis triggered by the decay process of
the $X^0$ particle might change the abundances of normal nuclei if the
energy injection by the decay were large enough.  Recent studies suggest
that the radiative decay could
lead to the production of $^6$Li to the level at most $\sim 10$ times larger
than that observed in MPHSs when the decay life is of the order of $\sim 10^8 -
10^{12}$~s which is associated with $^3$He
production~\cite{Kusakabe:2006hc}.  The hadronic decay, on the other
hand, can be a solution of
both the lithium problems although that case gives somewhat elevated
deuterium abundances~\cite{Jedamzik:2004er,Cumberbatch:2007me}.

For example, we assume that the mass and the initial abundance of the
$X$ are $m_X=100$~GeV and $Y_X=1.5\times 10^{-14}$, respectively.  The
energy injection at the decay of the $X$ into hadronic jets is
constrained to be $\lesssim O$(1--100 GeV) if the lifetime is $\tau_X
\gtrsim 10^2$~s through abundances of D, $^4$He, $^6$Li and $^3$He
depending upon the lifetime (figure 38 of
Ref.~\cite{Kawasaki:2004qu}).  The energy injection into
electromagnetic particles is, on the other hand, constrained to be
$\lesssim O$(10 GeV) if the lifetime is $\tau_X\gtrsim 10^7$~s (figure
1 of Ref.~\cite{Kusakabe:2006hc}).  This
amount of energy injection tends to attain $^6$Li production up to the
observed level in MPHSs.

We summarize a present status of several models which have effects and
thus leave observational signatures on primordial light element abundances.  In the BBN model catalyzed by a long-lived sub-SIMP studied in this paper, the abundance of $^7$Li can
be reduced below the level of SBBN prediction.  In the BBN model catalyzed by a long-lived SIMP, the abundances of $^9$Be or B can
be high~\cite{Kusakabe:2009jt}.  Moreover, the isotopic abundance ratio,
i.e., $^{10}$B/$^{11}$B can be high, which is never predicted in other
scenarios for boron production~\cite{Kusakabe:2009jt}.  In the BBN model
catalyzed by a negatively charged massive particle (CHAMP), the $^6$Li abundance
can be high~\cite{Pospelov:2006sc}.  Only if the abundance of the CHAMP is
more than $(0.04-1)$ times as large as that of baryon~\cite{Kusakabe:2010cb,Kusakabe2010inpc}, the
$^7$Li reduction can be possible~\cite{Bird:2007ge}.  A signature of CHAMP
on $^9$Be abundance has been estimated to be
negligible~\cite{Kusakabe:2010cb,Kusakabe2010inpc} in the light of a rigorous quantum
mechanical investigation~\cite{Kamimura2010}.  The cosmological
cosmic ray nucleosynthesis triggered by supernova explosion in an early
epoch of the structure formation can produce $^6$Li~\cite{Rollinde:2006zx}
as well as $^9$Be and $^{10,11}$B~\cite{Kusakabe2008,Rollinde2008}.  In baryon-inhomogeneous BBN models, the abundance of $^9$Be
can be higher than in the SBBN~\cite{Boyd1989,Kajino1990a,Kajino1990b,Coc1993,Orito1997}.

\appendix*

\section{Constraints on primordial light element abundances}\label{appendix}

We adopt constraints on primordial abundances as follows:

A mean value for the primordial deuterium abundance in Lyman-$\alpha$ absorption systems in the foreground of high redshift
QSOs has been estimated to be log(D/H)=$-4.55\pm 0.03$~\cite{Pettini:2008mq}.  We
adopt this value and a $2\sigma$ uncertainty, i.e.,
\begin{equation}
2.45\times10^{-5}< {\rm D}/{\rm H}< 3.24\times10^{-5}.
 \label{eq25}
\end{equation}

$^3$He abundances are measured in Galactic HII regions through the 8.665~GHz
hyperfine transition of $^3$He$^+$, i.e., $^3$He/H=$(1.9\pm 0.6)\times
10^{-5}$~\cite{Bania:2002yj}.  However, abundances in
extragalactic objects have not been measured, and it is not known whether $^3$He has increased or
decreased through the course of stellar and galactic evolution~\cite{Chiappini:2002hd,VangioniFlam:2002sa}.
$^3$He is more resistant to the stellar burning than deuterium.  Because
the deuterium abundance does not appear to have decreased since the
BBN epoch until the solar system formation~\cite{Lodders2003}, we do not
assume a decrease in the $^3$He abundance after BBN in amounts larger than the uncertainty in
the abundance determination of Galactic HII regions.  Although
a constraint on the primordial $^3$He abundance is rather weak considering
its uncertainty, we take a $2\sigma$ upper limit from abundances
in Galactic HII region as a rough guide, i.e.,
\begin{equation}
^3{\rm He}/{\rm H}< 3.1\times 10^{-5}.
 \label{eq26}
\end{equation}

For the primordial helium abundance we adopt two different constraints,
i.e, $Y=0.2565\pm 0.0051$~\cite{Izotov:2010ca} and $Y=0.2561\pm
0.0108$~\cite{Aver:2010wq} both from observations of metal-poor extragalactic
HII regions.  We take $2\sigma$ limits of 
\begin{equation}
0.2463 < Y < 0.2667~~~~~{\rm (IT10)},
\label{eq27}
\end{equation}
and
\begin{equation}
0.2345 < Y < 0.2777~~~~~{\rm (AOS10)}.
\label{eq28}
\end{equation}

An upper limit on the $^6$Li abundance is taken from the possible plateau abundance
of $^6$Li/H$=(7.1\pm 0.7)\times 10^{-12}$ observed in metal-poor halo
stars (MPHSs)~\cite{Asplund:2005yt}.  A $2\sigma$ uncertainty is included and we
derive
\begin{equation}
^6{\rm Li/H} < 8.5\times 10^{-12}.
\label{eq29}
\end{equation}

A limit on the $^7$Li abundance is taken from observations of
MPHSs, i.e., $^7$Li/H$=(1.23^{+0.68}_{-0.32})\times 10^{-10}$ (95\% confidence limits)~\cite{Ryan2000}.
The adopted constraint on the $^7$Li abundance is then
\begin{equation}
0.91\times 10^{-10} < {\rm ^7Li/H} < 1.91\times 10^{-10}.
\label{eq30}
\end{equation}

Although we do not use constraints on abundances of nuclei with mass
number $A\geq
9$ in this study, a primordial $^9$Be may be related to the scenario
(see Section~\ref{sec32}).  An upper limit on $^9$Be abundance should be
taken from the minimum abundances observed in MPHSs~\cite{Ito:2009uv}, i.e.,
\begin{equation}
^9{\rm Be/H} < 10^{-14}.
\label{eq31}
\end{equation}

\begin{acknowledgments}
This work is supported by Grant-in-Aid for JSPS
Fellows No.21.6817 (Kusakabe) and Grant-in-Aid for Scientific Research from the
Ministry of Education, Science, Sports, and Culture (MEXT), Japan,
No.22540267 and No.21111006 (Kawasaki) and also by World Premier International
Research Center Initiative (WPI Initiative), MEXT, Japan.
\end{acknowledgments}

\bibliography{reference}


\end{document}